\documentclass[aps,prx
,reprint,superscriptaddress,nofootinbib]{revtex4-2}
\usepackage{graphicx}
\usepackage{eucal}
\usepackage{import}
\usepackage{color,verbatim,bbold, float}
\usepackage{amssymb,amsbsy,mathrsfs}
\usepackage{times}
\usepackage{color}
\usepackage{placeins}
\usepackage{placeins}

\usepackage{setspace}
\usepackage[utf8]{inputenc}
\usepackage{bm}
\usepackage{braket}
\usepackage{commath}
\usepackage[dvipsnames]{xcolor}
\usepackage{csquotes}
\usepackage{enumitem}
\usepackage{dsfont}
\usepackage{appendix}
\usepackage{xcolor}
\usepackage{comment}
\usepackage[T1]{fontenc}
\usepackage{mathtools}
\usepackage{calc}
\usepackage[colorlinks,allcolors=blue]{hyperref}
\usepackage{physics}
\usepackage{tikz}
\usepackage{quantikz}

\begin{document}

\title{Bosonic quantum error-correcting codes with finite stellar rank}

\author{Rui Wang}
\affiliation{Department of Microtechnology and Nanoscience, Chalmers University of Technology, Göteborg SE-412 96, Sweden}
\author{Adithi Udupa}
\affiliation{Department of Microtechnology and Nanoscience, Chalmers University of Technology, Göteborg SE-412 96, Sweden}
\author{Timo Hillmann}
\affiliation{Department of Microtechnology and Nanoscience, Chalmers University of Technology, Göteborg SE-412 96, Sweden}
\affiliation{School of Physics, The University of Sydney, Sydney, New South Wales 2006, Australia}
\author{Ulysse Chabaud}
\affiliation{DIENS, \'Ecole Normale Sup\'erieure, PSL University, CNRS, INRIA, 45 rue d’Ulm, Paris, 75005, France}
\author{Alessandro Ferraro}
\affiliation{Dipartimento di Fisica “Aldo Pontremoli,” Università degli Studi di Milano, I-20133 Milano, Italy}
\author{Giulia Ferrini}
\affiliation{Department of Microtechnology and Nanoscience, Chalmers University of Technology, Göteborg SE-412 96, Sweden}

\begin{abstract}
Bosonic quantum error correction (QEC) relies on non-Gaussian bosonic encodings whose preparation cost is a central practical constraint.
In this work, we use stellar rank as a resource measure to design and benchmark bosonic codes under finite non-Gaussian resources.
For fixed cat and Gottesman--Kitaev--Preskill (GKP) code families, we show that finite stellar rank creates a trade-off among state approximability, energy, and logical protection under photon loss and photon-number dephasing, evaluated with optimal recovery. This trade-off implies that codewords with better ideal error-correction properties need not be optimal once finite-rank preparation constraints are imposed.
Going beyond fixed-target codewords, we directly optimize bosonic encodings at fixed stellar rank, revealing noise-adapted code structures and concrete resource thresholds. Grid-like encodings emerge under photon loss, whereas approximately rotation-symmetric encodings arise under dephasing.
In the optimized search, stellar rank $k=2$ suffices to surpass break-even for all dephasing strengths considered, while under photon loss the required rank increases with the loss rate. 
These results establish stellar rank as an operationally meaningful resource measure for bosonic QEC under practical state-preparation constraints.
\end{abstract}

\maketitle

\section{Introduction}
Bosonic codes provide a promising route to quantum error correction by encoding logical qubits into the infinite-dimensional Hilbert space of a bosonic mode~\cite{Joshi2021IOP, Terhal2015RMP}.
Compared with conventional discrete-variable QEC schemes, which use redundancy across many physical qubits~\cite{Shor1995PRA, Steane1996PRL}, bosonic codes exploit continuous-variable oscillator states to suppress dominant noise processes such as photon loss and dephasing while improving hardware efficiency~\cite{Chuang1997PRA, Cochrane1999PRA, Michael2016PRX}.
Examples include cat codes~\cite{Nokkala2018NJP, Bergmann2016PRA}, binomial codes~\cite{Michael2016PRX}, and Gottesman--Kitaev--Preskill (GKP) codes~\cite{Gottesman2001PRA, Baragiola2019PRL}, which have seen substantial experimental progress~\cite{gkpxanadu2025Nature, Putterman2025Nature, Ni2023Nature}.

In bosonic platforms, Gaussian states and operations constitute the most readily available experimental toolbox, since they are generated by Hamiltonians which are quadratic in the canonical operators, and can be implemented using linear optics and parametric squeezing processes \cite{Ferraro05, Weedbrook2012RMP, Brask2022arxiv}. 
However, bosonic QEC encodings, such as cat and GKP states, are intrinsically non-Gaussian, making non-Gaussian resources essential for their preparation and implementation \cite{Niset2009PRL}.
In superconducting microwave platforms, non-Gaussian oscillator states can be generated by combining displacement operations with selective number-dependent arbitrary phase (SNAP) gates \cite{Heeres2015PRL, Krastanov2015PRA, Kudra2022PRX.Q}, as well as through native cubic interactions \cite{hillmann2020universal, eriksson2024universal} or quantum lattice gates \cite{Lingzhen2025CP, huang2026arxiv}.
In trapped-ion systems, analogous states can be engineered through coherent control of the ion’s internal and motional degrees of freedom using laser pulses, as well as through intrinsic or effective nonlinearities such as Kerr-type interactions \cite{kirchmair2013nature, Leibfried2003RMP, Bazavan2026np}. 
These approaches provide flexible and high-fidelity control, making superconducting cavities and trapped ions leading platforms for near-term demonstrations of bosonic codes.

By contrast, photonic systems, which offer the advantage of room temperature and on-chip implementations \cite{Aghaee2025Nature, zhong2020Science, madsen2022nature} as well as  scalable entanglement generation \cite{Yoshikawa2016APL, Yokoyama2013NP} only provide limited optical nonlinearities, making the preparation of non-Gaussian codewords substantially more challenging \cite{Braunstein2005RMP}.
In practice, experimentally accessible non-Gaussian operations are often based on photon subtraction or photon addition \cite{Barnett2018PRA, Landazabal2025NJP, Yokoyama2013NP}, possibly interleaved with readily available Gaussian operations such as displacements and squeezing. 
This motivates a quantitative understanding of the trade-offs between non-Gaussian resource cost, state-preparation accuracy, and the resulting error-correction performance.

To quantify these resource requirements, particularly in photonic settings, recent theoretical work has introduced the notion of stellar rank, which characterizes the minimal number of photon additions required to generate a given single-mode non-Gaussian state \cite{ulysse2020PRL}.
Stellar rank thus provides a natural measure of the non-Gaussian resources needed to prepare bosonic codewords, directly connecting resource-theoretic descriptions to experimentally accessible operations such as photon addition and Gaussian unitaries.
Ideal non-Gaussian states can in practice only be prepared approximately under finite resource constraints. An appropriate measure of the quality of non-Gaussian approximated states is stellar fidelity, defined for a given stellar rank as the maximal achievable fidelity with a target state over all states of that stellar rank~\cite{Chabaud2021PRX.Q, Hahn2026quantum}.

These concepts are particularly relevant for realistic implementations, where non-Gaussian resources are inherently limited. Nevertheless, it remains unclear how finite-stellar-rank approximations constrain the achievable fidelity of bosonic codewords and, ultimately, their QEC performance.
In particular, it is an open problem to determine the minimal stellar rank needed to achieve a practical error correction advantage, i.e., to surpass the break-even point ~\cite{Ofek2016Nature,hu2019NP}, defined by the performance of an unencoded qubit subject to the same noise.

In this work, we develop a resource-constrained framework for analyzing and designing non-Gaussian bosonic codewords under finite stellar-rank constraints. 
We first study how accurately standard bosonic code families, including cat and Gottesman--Kitaev--Preskill (GKP) codes, can be approximated with bounded stellar rank, and quantify the achievable approximation quality using stellar fidelity \cite{Chabaud2021PRX.Q,Hahn2026quantum}.
We then connect these state-preparation constraints to operational quantum error correction performance by benchmarking the resulting approximate codewords under physically relevant noise channels, including photon loss and dephasing, using optimal recovery \cite{Kosut2009QIP}. 
This shows that high stellar fidelity alone does not guarantee good logical performance: codewords that are favorable in the ideal infinite-resource limit may become suboptimal when only finite stellar rank is available.

We then go beyond fixed code families and optimize bosonic encodings directly within finite-stellar-rank variational classes. For computational tractability, we use the Petz-recovery fidelity as a surrogate objective for finding candidate encodings, and benchmark the final encodings using optimal recovery. This reveals noise-adapted code structures and quantitative resource thresholds for surpassing break-even, extending previous noise-adapted encoding studies to the practically relevant setting of constrained non-Gaussian state preparation \cite{Leviant2022quantum}.

The paper is organized as follows.
Section~\ref{sec:basics} reviews the general framework of bosonic QEC, including bosonic codes, noise channels, recovery channels and the definitions of stellar rank and stellar fidelity used to characterize finite-resource approximations of bosonic codewords.
Section~\ref{sec:QEC_approximate codewords} presents analytic constructions of cat-code and GKP-code approximations under finite stellar-rank constraints, and benchmarks their performance under photon-loss and dephasing noise using optimal recovery.
Section~\ref{sec:QEC_optimal_codewords} goes beyond fixed code families by optimizing encodings directly at fixed stellar rank, using the Petz-recovery fidelity as a search objective and optimal recovery as the final performance benchmark.
Finally, Section~\ref{sec:conclusion} concludes the paper and discusses the implications for resource-limited bosonic QEC.

\section{Background} 
\label{sec:basics}

In this Section, we review the theoretical framework of bosonic error correction and introduce stellar rank as a measure of non-Gaussianity.

\subsection{Bosonic codes}

A convenient way to characterize bosonic codes is through the symmetry structure of their code space in phase space.
In this work, we focus on two representative symmetry classes: rotation-symmetric bosonic (RSB) codes \cite{Arne2020PRX}, whose code spaces are invariant under discrete phase-space rotations, and codes with translational symmetry, namely GKP codes \cite{Gottesman2001PRA}, whose structure is defined by periodic displacements in phase space.
We briefly review these typologies below. 

\subsubsection{Rotation-symmetric bosonic codes}
RSB codes are defined by an $N$-fold discrete rotational symmetry in phase space.
A bosonic encoding defines a two-dimensional logical subspace with projector,
\begin{equation}
\hat{\Pi}_L = |0_L\rangle\langle 0_L| + |1_L\rangle\langle 1_L|,
\end{equation}
Here and throughout, the codewords refer to an orthogonal and normalized logical basis $\{\ket{0_L},\ket{1_L}\}$, while the code denotes the subspace projected onto by $\hat{\Pi}_L$. The encoding has an $N$-fold rotational symmetry if $\hat{\Pi}_L$ commutes with the rotation operator
\begin{equation}
\hat{R}_N = e^{i \tfrac{2\pi}{N} \hat{n}},
\end{equation}
where $\hat{n} = \hat{a}^\dagger \hat{a}$ is the number operator.

An order-$N$ RSB code further requires that
\begin{equation}
\hat{Z}_N = \hat{R}_{2N} = e^{i\pi \hat{n}/N}
\end{equation}
acts as the logical $\hat{Z}$ operator with eigenvalues $\pm 1$ on the code space \cite{Arne2020PRX, Albert2018PRA}.
Under this condition, the logical states can be constructed as superpositions of rotated versions of a primitive state $\ket{\Theta}$,
\begin{align}
|\mu_L\rangle &= \frac{1}{\sqrt{\mathcal{N}_\mu}} \sum_{z=0}^{2N-1} (-1)^{\mu z}e^{i\frac{z\pi}{N}\hat{n}} |\Theta\rangle, \qquad (\mu=0, 1),
\end{align}
where $\mathcal{N}_{\mu}$ is a codeword dependent normalization constant.
 
A prominent example is the cat code, obtained by choosing $\ket{\Theta}=\ket{\alpha}$ to be a coherent state \cite{Cochrane1999PRA, Mirrahimi2014NJP}.  
For a general cat state with arbitrary rotational symmetry $N$ and coherent amplitude $\alpha$, the logical basis states can be written compactly as 
\begin{align}
|\mathrm{Cat}_{N,\alpha}^\mu\rangle
    &= \frac{1}{\sqrt{\mathcal{N}_{\mu}}} 
       \sum_{z=0}^{2N-1} (-1)^{\mu z } |e^{i z \pi/N}\alpha\rangle.
       \label{eq:def_general_cat_codes}
\end{align}

The logical states have support on disjoint Fock subspaces distinguished by photon number modulo $2N$.
Photon-loss errors therefore move the codewords between different rotational-symmetry sectors, a feature that underlies the error-correcting properties of RSB codes \cite{Ofek2016Nature,Leghtas2015Science,Arne2020PRX}.

\subsubsection{Gottesman--Kitaev--Preskill codes}

Another important class of bosonic codes is given by the Gottesman--Kitaev--Preskill (GKP) code, which exhibits discrete translational symmetry in phase space~\cite{Gottesman2001PRA}. Throughout this work, we focus on the square-lattice GKP code, defined by the stabilizers
\begin{equation}
\begin{split}
    &\hat{S}_X = e^{-i2\sqrt{\pi}\hat{p}}, \\
&\hat{S}_Z = e^{i2\sqrt{\pi}\hat{x}},
\end{split}
\label{eq:GKP_stabilizers}
\end{equation}
which enforce a periodic square-grid structure in phase space.
In the position basis, the corresponding ideal (non-normalizable) logical codewords can be written as
\begin{align}
    \ket{\mathrm{GKP}^{\mu}} &\propto \sum_{n\in\mathbb{Z}} \ket{(2n+\mu)\sqrt{\pi}}_x,
\end{align}
where $\ket{x}_x$ is a generalized eigenstate of the position operator.
This grid structure enables the correction of small displacement errors in both quadratures, making GKP codes naturally suited for Gaussian noise and small shift errors.

Equivalently, the ideal square-lattice GKP codewords can be represented in the coherent-state basis~\cite{Gottesman2001PRA, Grimsmo2021PRX.Q} as
\begin{equation}
    \ket{\mathrm{GKP}^{\mu}}
    \propto
    \sum_{g,l\in\mathbb{Z}}
    e^{-i\pi l(g+\mu/2)}
    |
    \sqrt{\frac{\pi}{2}}
    \left(2g+il+\mu\right)
    \rangle,
\end{equation}
where $\ket{\zeta}=\hat{D}(\zeta)\ket{0}$ is a coherent state. 
This representation is particularly convenient for the calculation of stellar fidelity presented in Sec.~\ref{sec:stellar_fidelity_GKP}.

However, ideal GKP states are unphysical; they are non-normalizable and have infinite energy.
In practice, approximate GKP states are obtained by applying an energy-damping envelope~\cite{Menicucci2014PRL,Royer2020PRL}.
Finite-energy GKP states and their error-correction properties have been extensively studied in both theoretical and experimental settings \cite{Glancy2006PRA, Fukui2018PRX, Albert2018PRA}.

We express finite-energy codewords through the damping operator as
\begin{equation}
\ket{\mathrm{GKP}^{\mu}_\Delta}
    \propto
    e^{-\Delta^2\hat{n}}
    \ket{\mathrm{GKP}^{\mu}},
\end{equation}
which gives
\begin{align}
\ket{\mathrm{GKP}^{\mu}_\Delta}
    &=
    \frac{1}{\sqrt{\mathcal{N}_\mu}}
    \sum_{g,l\in\mathbb{Z}}
    e^{-i\pi l(g+\mu/2)}
    e^{-\Delta^2|\alpha_{gl}|^2}
    |e^{-\Delta^2}\alpha_{gl}\rangle,
    \label{eq:codewords-GKP}
\end{align}
where $\mathcal{N}_\mu$ is a normalizing factor and with
\begin{equation}
    \alpha_{gl}=
    \sqrt{\frac{\pi}{2}}(2g+il+\mu).
\end{equation}
The damping parameter $\Delta$ controls the finite-energy approximation: smaller $\Delta$ corresponds to a less damped, higher-energy state closer to the ideal grid limit, while larger $\Delta$ gives a more strongly damped, lower-energy state.

\subsection{Noise and recovery channel}
\label{sec:noise_recovery}
Having introduced the structure and symmetry properties of bosonic codewords, we now turn to their performance under physical noise and recovery processes.
A bosonic quantum error-correcting (QEC) protocol consists of three stages: encoding, noise, and recovery.
Throughout this work, we focus on encoding a single logical qubit, described by a density operator $\hat{\rho}_S$, on the two-dimensional Hilbert space $\mathcal{H}_S \simeq \mathbb{C}^2$, into the bosonic Hilbert space $\mathcal{H}_C$ through an encoding map $\mathcal{E}$.
The encoded state then evolves under a physical noise channel $\mathcal{N}$, followed by a recovery operation $\mathcal{R}$ \cite{Kosut2009QIP,Nielsen2010CUP}.
The resulting effective logical channel is
\begin{equation}
    \mathcal{C}
    =
    \mathcal{R}\circ\mathcal{N}\circ\mathcal{E}.
\end{equation}
A detailed Kraus operator formulation of the encoding, noise, and recovery maps is provided in Appendix~\ref{appendix:qec-framework}.

In this work, we focus on two physically relevant bosonic noise channels: photon loss and dephasing, which are widely used to model dominant errors in oscillator-based quantum memories \cite{Michael2016PRX,Albert2018PRA}.
Photon loss describes energy relaxation due to leakage of excitations from the oscillator and is characterized by the loss rate $\kappa$.
Dephasing captures phase randomization without energy decay and is characterized by the dephasing rate $\kappa_\phi$.
Both noise channels are represented in Kraus form.
Their explicit forms are summarized in Appendix~\ref{appendix:photon-loss} and Appendix~\ref{appendix:dephasing}.

To quantify the performance of the effective logical channel $\mathcal{C}$, we use the channel fidelity \cite{Fletcher2007PRA,Schumacher1996PRA}.
This quantity measures how well the channel preserves logical information, including correlations with an ideal reference system. For a channel $\mathcal{C}$ acting on a logical qubit, the channel fidelity is defined as
\begin{align}
    F(\mathcal{C})
    &=
    \bra{\Gamma}
    \hat{\rho}_{\mathcal{C}}
    \ket{\Gamma}
    \nonumber\\
    &=
    \frac{1}{4}
    \sum_{\mu,\nu=0}^1
    \bra{\mu}
    \mathcal{C}(\ketbra{\mu}{\nu})
    \ket{\nu},
    \label{eq:entanglement-fidelity}
\end{align}
where
\begin{equation}
    \hat{\rho}_{\mathcal{C}}
    =
    (\mathcal{C}\otimes I)
    (\ketbra{\Gamma}{\Gamma}),
    \qquad
    \ket{\Gamma}
    =
    \frac{1}{\sqrt{2}}
    \sum_{\mu=0}^{1} \ket{\mu\mu}.
\end{equation}
Here, $(\mu,\nu\in{0,1})$ label the two logical computational-basis states spanning the system Hilbert space $\mathcal{H}_{S}$.
For a fixed encoding $\mathcal{E}$ and noise channel $\mathcal{N}$, the optimal recovery is obtained by maximizing the channel fidelity over all recovery maps \cite{Kosut2009QIP,Fletcher2007PRA},
\begin{equation}
   \mathcal{R}^{\mathrm{opt}}
    =
    \underset{\mathcal{R}}{\mathrm{argmax}}
    \;
    F(
    \mathcal{R}
    \circ
    \mathcal{N}
    \circ
    \mathcal{E}
    ).
\end{equation}

\subsection{Stellar rank and stellar fidelity}
\label{sec:stellar_profile}
The practical utility of bosonic codes relies crucially on the ability to prepare their codeword states with sufficient accuracy.
Both cat and GKP codewords are highly non-Gaussian, and 
their preparations therefore requires resources beyond Gaussian operations. Such non-Gaussian resources are widely regarded as essential for achieving quantum computational advantage in continuous-variable architectures \cite{Mari2012PRL, Lloyd1999PRL}.

From a resource-theoretic perspective, the non-Gaussianity of a single-mode pure state $\ket{\psi}$ can be quantified by its stellar rank $r^\star(\psi)$, 
defined as the number of zeros, counted with half multiplicity, of its Husimi $Q$-function,
\begin{equation}
Q_\psi(\alpha)
=
\frac{1}{\pi}
\abs{\braket{\alpha}{\psi}}^2.
\end{equation}
Equivalently, the stellar rank is the minimal number of photon additions required to generate the state when arbitrary Gaussian operations are available \cite{ulysse2020PRL}.
For a single-mode pure state with finite stellar rank, one can write
\begin{equation}
    \ket{\psi} = \frac{1}{\sqrt{\mathcal{N}_0}} 
    \Bigg[ \prod_{n=1}^{r^\star(\psi)} 
    \hat{D}(\alpha_n)\,\hat{a}^\dagger\,\hat{D}^\dagger(\alpha_n) 
    \Bigg] \ket{\psi_G},
\end{equation}
where $\ket{\psi_G}$is a Gaussian state, $\hat{D}
(\alpha_n)$ is the displacement operator, the parameters $\{\alpha_n\}$ are the roots of $Q_\psi(\alpha)$, counted with half multiplicity, and $\mathcal{N}_0$ is a normalization constant.

In practical settings, however, preparing an exact target state is typically infeasible due to limited experimental resources, particularly for highly non-Gaussian states requiring many non-Gaussian operations.
This motivates the notion of approximate stellar rank, which quantifies the minimal number of photon additions required to approximate a given non-Gaussian state up to a specified accuracy \cite{Hahn2026quantum}:
\begin{equation}
    r_\epsilon^\star(\psi) = \inf_{\ket{\phi}: \, F(\psi,\phi) \geq 1-\epsilon} \, r^\star(\phi) ,
\end{equation}
where $F(\psi,\phi)$ denotes the fidelity between $\ket{\psi}$ with $\ket{\phi}$
In other words, $r_\epsilon^\star(\psi)$ is the smallest stellar rank such that $\ket{\psi}$ can be approximated within fidelity $1-\epsilon$.
In the limit $\epsilon \to 0$, one recovers the exact stellar rank, $r_\epsilon^\star(\psi) \to r^\star(\psi)$.

Complementary to this notion, one can ask how well a given target state can be approximated using states of bounded stellar rank.
This is captured by the stellar fidelity, defined as
\begin{equation}
    f_k^\star(\psi) = \max_{\phi:\, r^\star(\phi)\leq k}F(\psi,\phi), 
    \qquad k \in \mathbb{N}.
\end{equation}
For single-mode pure states, the stellar fidelity admits an efficient formulation as an optimization over Gaussian unitaries~\cite{Chabaud2021PRX.Q},
\begin{equation}
    f^\star_k(\psi)
    =
    \max_{\hat{G}\in\mathcal{G}}
    \bra{\psi}\hat{G}^\dagger \hat{\Pi}_{k}\hat{G}\ket{\psi},
    \label{eq:def_stellar_fidelity}
\end{equation}
where the maximization is performed over the set $\mathcal{G}$ of single-mode Gaussian unitaries, and
\begin{equation}
    \hat{\Pi}_k
    =
    \sum_{m=0}^{k}\ket{m}\bra{m}
\end{equation}
is the projector onto the finite Fock subspace spanned by
$\{\ket{0},\ldots,\ket{k}\}$.
We parametrize the Gaussian unitary as
\begin{equation}
    \hat{G}(\zeta,\beta)
    =
    \hat{S}(\zeta)\hat{D}(\beta),
\end{equation}
where $\hat{S}(\zeta)$ and $\hat{D}(\beta)$ denote the squeezing and displacement operators, respectively.
The optimization is therefore performed over the squeezing parameter $\zeta$ and displacement parameter $\beta$.
Once the optimal Gaussian unitary $\hat{G}_0=\hat{G}(\zeta_0,\beta_0)$ is found, the optimal stellar-rank-$k$ approximating state is constructed as~\cite{Chabaud2021PRX.Q}
\begin{equation}
    \ket{\psi_{\mathrm{opt}}}
    =
    \hat{G}_0^\dagger
    \left(
    \frac{\hat{\Pi}_k \hat{G}_0\ket{\psi}}
    {\left\|\hat{\Pi}_k \hat{G}_0\ket{\psi}\right\|}
    \right).
\end{equation}

Together, approximate stellar rank and stellar fidelity provide complementary perspectives on the cost–accuracy trade-off in preparing non-Gaussian bosonic codewords, and will serve as central tools in the analysis that follows.

\section{QEC with approximate codewords at finite stellar rank}
\label{sec:QEC_approximate codewords}
We investigate how finite stellar rank constrains the approximability of standard bosonic codewords and their error-correction performance.
We first analyze cat and GKP code states via stellar fidelity to obtain optimized finite-stellar-rank approximations, and then benchmark the resulting approximate encodings under photon-loss and dephasing noise with optimal recovery.

For both cat and GKP stellar profiles,  the stellar-fidelity optimization is performed over a non-convex parameter space.
Therefore, the numerical values reported here should be interpreted as optimized lower bounds rather than certified global optima.
This limitation becomes more pronounced for highly non-Gaussian target states, where the optimization landscape is more complex, and the final result can depend on the initial point and random seed.
In practice, this effect will be visible for larger-amplitude cat states and less regularized GKP states with smaller $\Delta$.
Consequently, the optimized stellar fidelities may not appear strictly monotonic with stellar rank, even though the true stellar fidelity is increasing by definition.

\subsection{Stellar profiles of cat codes and GKP codes}
\label{sec:approximate_codeword_preparation}

We begin by evaluating and analyzing the stellar fidelity of cat and GKP states using Eq.~(\ref{eq:def_stellar_fidelity}) applied to the case of these target states, and analyzing the corresponding stellar fidelities.

\subsubsection{Stellar fidelities of cat codes}
\begin{figure*}[t]
    \centering
    \includegraphics[width=\textwidth]{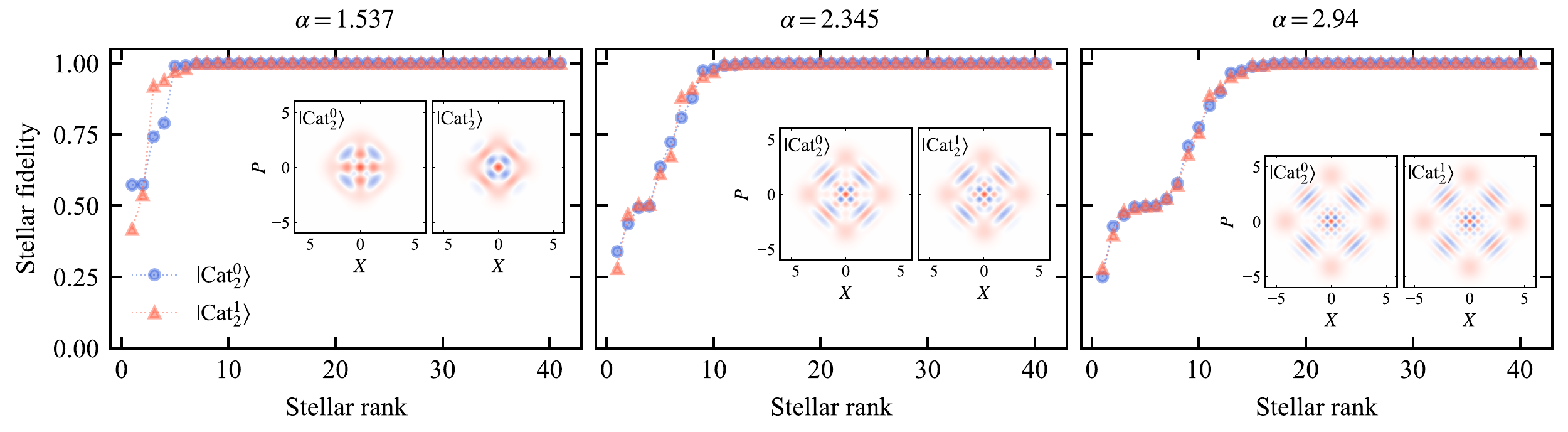}
    \caption{Stellar fidelity as a function of stellar rank $k$ for the logical $ |\mathrm{Cat}_{2}^0\rangle$ (circles) and $ |\mathrm{Cat}_{2}^1\rangle$ (triangles) states of cat codes with rotational symmetry order $N=2$.
    Results are shown for three cat-state amplitudes at the sweet spots $\alpha = 1.537$, $\alpha = 2.345$, and $\alpha = 2.94$.
    For larger $\alpha$, the convergence with stellar rank becomes slower and plateau-like features appear, reflecting the increasing non-Gaussian resources required to resolve more separated coherent-state components.
    The insets show the Wigner functions of the target states being approximated.
    A detailed discussion of the plateau behavior is provided in Appendix~\ref{appendix:fidelity_platuea_cat}.}
\label{fig:stellar_profile_wignerfunction_cat}
\end{figure*}

For the  cat code vectors $|\mathrm{Cat}_{N,\alpha}^{\mu}\rangle$  in Eq.~(\ref{eq:def_general_cat_codes}), the stellar fidelity becomes
\begin{align}
   &f^\star_k(|\mathrm{Cat}_{N,\alpha}^{\mu}\rangle) =
   \max_{\hat{G}\in\mathcal{G}}
    \sum_{m=0}^{k}
    \left|
    \left\langle m\right|
    \hat{G}
    |\mathrm{Cat}_{N,\alpha}^{\mu}\rangle
    \right|^2\nonumber\\
    & = \max_{\hat{G}\in \mathcal{G}} \frac{1}{\sqrt{\mathcal{N_{\mu}}}} \sum_{m=0}^k \sum_{z=0}^{2N-1} \Big|(-1)^{\mu z}\bra{m} \hat{G} |\tilde{\alpha}\rangle\Big|^2,
\end{align}
where we set $\tilde{\alpha}=e^{i z \pi/N}\alpha$.

Using Eq.~(\ref{eq:def_stellar_fidelity}) we obtain the stellar fidelity of general cat states as
\begin{align}
&f^\star_k(|\mathrm{Cat}_{N,\alpha}^{\mu}\rangle)\nonumber\\
    &= \max_{\hat{G}\in \mathcal{G}}
       \frac{e^{-(|\alpha|^2+|\beta|^2)}}{\mathcal{N}_{\mu} \cdot c_r}
       \sum_{m=0}^k \frac{t_r^m}{m!}
       \Big\lvert
        \sum_{z=0}^{2N-1}
        (-1)^{\mu z}
        u_m(\tilde{\alpha},\zeta,\beta)
        \Big\rvert^2,
       \label{eq:stellar_fidelity_cat}
\end{align}
where
\begin{equation}
u_m(\tilde{\alpha}, \zeta, \beta) =
e^{-\tilde{\alpha}\beta^\star+\tfrac{1}{2}e^{i\theta}t_r(\tilde{\alpha}+\beta)^2}
{\rm {H}e}_m\left(\tfrac{\tilde{\alpha}+\beta}{\sqrt{s_rc_r}}e^{i\theta/2}\right),
\end{equation}
with $\zeta = r e^{i\theta}$ 
and
\begin{equation}
    c_r = \cosh r,\quad
s_r = \sinh r,\quad
t_r = \tanh r. \nonumber
\end{equation}
Here, ${\rm {H}e}_m$ are the Hermite polynomials
\begin{align}
{\rm {H}e}_m(x)
&=(-1)^{m}e^{x^{2}/2}
\frac{\mathrm{d}^m}{\mathrm{d}x^m}
e^{-x^2/2}\nonumber\\
&=\sum_{p=0}^{\left\lfloor m/2\right\rfloor}
\frac{m!(-1)^{p}x^{m-2p}
}{2^{p}p!(m-2p)!}.
\end{align}
for all $m\in \mathbb{N}$ and all $x\in \mathbb{C}$.
We give a full derivation of Eq.\eqref{eq:stellar_fidelity_cat} in Appendix~\ref{appendix:stellar_fidelity_cat}, which extends previous results obtained for the case of two-component cat states \cite{Chabaud2021arXiv}. 

Using Eq.~(\ref{eq:stellar_fidelity_cat}), we numerically optimize the stellar fidelity for each target cat codeword over Gaussian unitaries, thereby identifying the best bounded-rank approximation within the chosen search space.
The Gaussian unitary in Eq.~(\ref{eq:def_stellar_fidelity}) is parameterized by displacement and squeezing operations, with $\lvert\beta\rvert\in[0,3]$, squeezing strength $r\in[0,3]$, squeezing phase $\theta\in[0,2\pi]$, and $\zeta=re^{i\theta}$.
The optimization is performed using the Broyden-Fletcher-Goldfarb-Shanno (BFGS) algorithm, as implemented in \texttt{scipy.optimize.minimize} \cite{Virtanen2020SciPy}. For each target codeword and stellar rank $k$, we use $10$ independent random initializations drawn uniformly from the parameter domain and retain the run with the highest stellar fidelity.
All $10$ random restarts converged to the same stellar fidelity within numerical precision for every code vector and stellar rank considered, which we take as numerical evidence that, in this case, the identified optimum coincides with the global maximum rather than a local one. Given that the optimization is non-convex, we additionally verified that enlarging the search window did not appreciably change the optimized stellar fidelity.

We compare the optimized stellar fidelities for cat codes with different coherent-state amplitudes $\alpha$, while fixing the rotational symmetry order to $N=2$.
The chosen amplitudes $\alpha$ correspond to known sweet-spot values for ideal QEC performance~\cite{Li2017PRL}. They therefore provide physically motivated target code vectors for assessing how finite stellar rank affects the preparation accuracy of cat codes.  The results are shown in Fig.~\ref{fig:stellar_profile_wignerfunction_cat}, where the insets display the Wigner function of the corresponding  target cat states. The observed trends are representative of target cat states with comparable amplitudes, and are not specific to the particular sweet-spot values chosen here.

From Fig.~\ref{fig:stellar_profile_wignerfunction_cat} we observe that the optimized stellar fidelity increases with the allowed stellar rank, reflecting the improved approximation quality provided by additional non-Gaussian resources.
However, the convergence strongly depends on the coherent amplitude $\alpha$.
For larger $\alpha$, the coherent components of the cat state become more widely separated in phase space, and higher stellar rank is required to resolve the full rotational structure accurately.
As a result, the approach to unit stellar fidelity becomes slower as $\alpha$ increases.
\begin{figure*}[t]
\centering
\includegraphics[width=\textwidth]{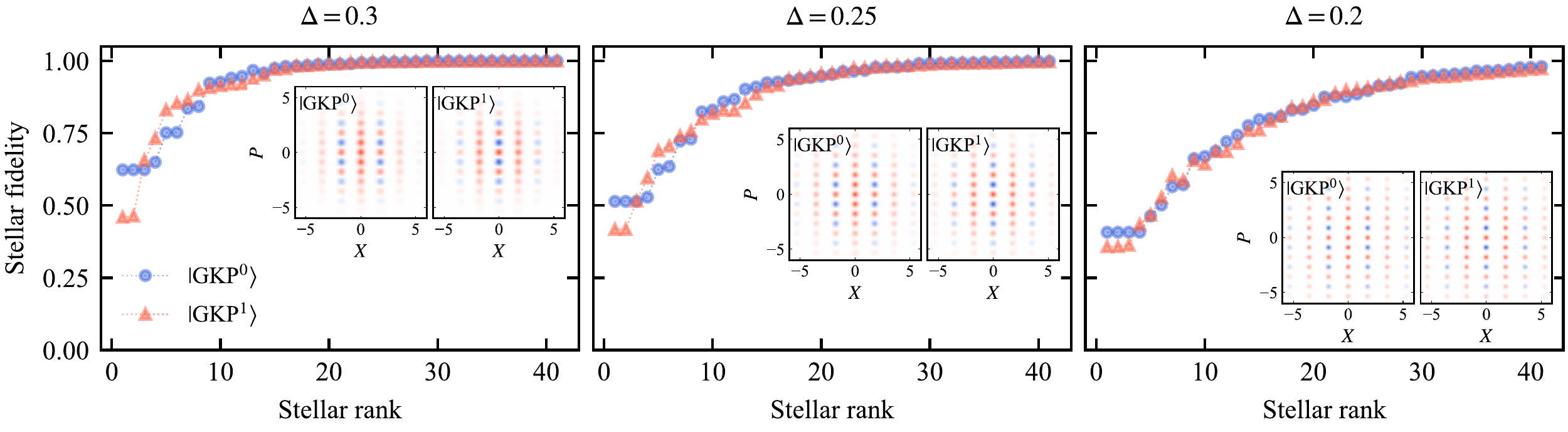}
    \caption{Approximation of GKP codewords with finite stellar rank.
    The optimized stellar fidelity is plotted versus stellar rank $k$ for the logical $|\mathrm{GKP}^{0}\rangle$ (circles) and $|\mathrm{GKP}^{1}\rangle$ (triangles) states with regularization parameters $\Delta=0.3$, $0.25$, and $0.2$.
    Smaller $\Delta$ produces sharper grid peaks and a more extended phase-space structure, requiring higher stellar rank to reach comparable fidelity.
    The insets show the Wigner functions of the target GKP states.}
\label{fig:stellar_profile_wignerfunction_gkp}
\end{figure*} 
A notable feature is the emergence of plateau regions, most prominently around stellar fidelity $f_k^\star \simeq 0.5$.
These plateaus indicate intermediate regimes in which the finite stellar rank approximation captures only part of the target cat-code structure before sufficient stellar rank is reached to resolve the remaining rotational components.
We further find that the number and structure of these plateaus are tied to the rotation symmetry order and amplitude of the codewords.
A detailed analysis of this plateau behavior is provided in Appendix~\ref{appendix:fidelity_platuea_cat}.

Having established the stellar fidelity profiles for cat codes, we now turn to GKP codes, whose extended grid structure in phase space presents a qualitatively different approximation challenge.

\subsubsection{Stellar fidelities of GKP codes}
\label{sec:stellar_fidelity_GKP}

Here, we derive the stellar fidelity of the approximate logical codeword states for GKP states given in Eq.~(\ref{eq:codewords-GKP}) by using the definition of stellar fidelity in Eq.~(\ref{eq:def_stellar_fidelity}); we obtain
\begin{align}
&f^\star_k(\ket{\text{GKP}^\mu_\Delta})\nonumber\\
    &=
    \max_{\hat G}
    \frac{1}{\mathcal N_\mu \cdot c_r}
    \sum_{m=0}^{k}
    \frac{t_r^m}{m!}
    \Big|
    \sum_{g,l\in \mathbb{Z}}
    e^{-i\pi l (g + \frac{\mu}{2})}
    e^{-\Delta^2|\alpha_{gl}|^2}
    \nonumber\\
    &
    \times e^{-\frac12(|\tilde{\alpha}_{gl}|^2+|\zeta|^2)-\tilde{\alpha}_{gl} \zeta^\star}
    e^{\frac12 e^{i\theta}t_r(\zeta+\tilde{\alpha}_{gl})^2}\nonumber\\
    & \times {\rm {H}e}_m\!\left(
    \frac{\zeta+\tilde{\alpha}_{gl}}
    {\sqrt{c_r\,s_r}}
    e^{i\theta/2}
    \right)
    \Big|^2,
    \label{eq:stellar_fidelity_gkp}
\end{align}
where $\tilde{\alpha}_{gl}=e^{-\Delta^2}\alpha_{gl}$.
The full derivation of Eq.~(\ref{eq:stellar_fidelity_gkp}) is given in Appendix~\ref{appendix:stellar_fidelity_GKP}. 
This expression allows us to optimize finite-stellar-rank approximations of energy-damped GKP codewords in the same way as for cat codes, using the same BFGS optimization procedure and random-restart strategy described above.

We compare the optimized stellar fidelities for approximate GKP states with different regularization parameters $\Delta=0.3$, $0.25$, and $0.2$, as shown in Fig.~\ref{fig:stellar_profile_wignerfunction_gkp}.
As $\Delta$ decreases, the GKP peaks become sharper and the grid extends over a large phase-space region, so higher stellar rank is required to reach comparable fidelity.
Accordingly, the more strongly regularized states with larger $\Delta$ are approximated more efficiently at fixed stellar rank.

For $\Delta=0.3$, both logical code vectors reach nearly unit stellar fidelity within the stellar-rank range considered here, $k \le40$.
In contrast, for smaller-$\Delta$ remains visibly harder to approximate: the optimized stellar fidelities reach only $0.981$ for the logical $|\mathrm{GKP}^{0}\rangle$ state and $0.973$ for the logical $|\mathrm{GKP}^{1}\rangle$ state within the range $k=40$. The $\Delta=0.2$ curves also show small non-monotonic features at intermediate stellar rank, consistent with the increased numerical difficulty of optimizing less regularized GKP targets. However, the overall trend remains that decreasing $\Delta$ increases the stellar-rank resources required for accurate preparation.

Note that for both multi-component cat states and GKP states, the stellar fidelities are evaluated using the analytically simplified expressions in Eqs.\eqref{eq:stellar_fidelity_cat} and \eqref{eq:stellar_fidelity_gkp}.
These expressions reduce the optimization to a search over  Gaussian parameters only, rather than requiring a brute-force optimization over truncated target states in the Fock basis.
This substantially lowers the computational cost, especially for highly non-Gaussian states with large phase-space support.
In principle, the same stellar-fidelity profiles could also be reproduced by a direct numerical optimization using the openly available library of Ref.~\cite{stellarnumerics2026}.

\subsection{Performance of approximate codewords}
\label{sec:qec_benchmark}

\begin{figure*}[t]
    \centering
    \includegraphics[width=\textwidth]{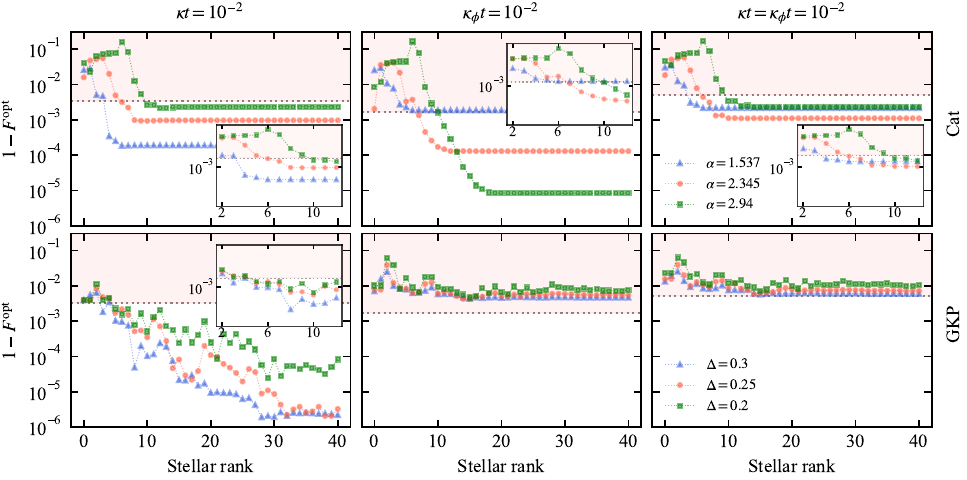}
    \caption{
    Error-correcting performance of finite-stellar-rank approximations of cat and GKP codewords.
    The optimal channel infidelity $1-F^{\mathrm{opt}}$ is plotted as a function of stellar rank $k$.
    The upper row shows results for the $N=2$ cat code with amplitudes $\alpha=1.537$ (blue), $\alpha=2.345$ (red), and $\alpha=2.94$ (green).
    The lower row shows results for approximate GKP codes with Gaussian damping parameters $\Delta=0.3$ (blue), $\Delta=0.25$ (red), and $\Delta=0.2$ (green).
    From left to right, the columns correspond to photon loss with $\kappa t=10^{-2}$, dephasing with $\kappa_\phi t=10^{-2}$, and combined noise with $\kappa t=\kappa_\phi t=10^{-2}$.
    Markers denote the optimal channel infidelity at certain stellar ranks, while the connecting dashed lines are guides to the eye. The horizontal threshold dashed lines indicate the break-even infidelity, and the shaded regions denote the regimes in which the finite-rank approximate codewords perform worse than the break-even threshold. The insets zoom in on representative stellar-rank intervals around the crossings of the break-even threshold; they are placed over regions where the corresponding main-panel curves have already saturated.}
        \label{fig:gate_average_infidelity_cat_gkp}
\end{figure*}

We now evaluate how the stellar-rank-limited state preparation analyzed above translates into logical error-correction performance. For each stellar rank $k$, we use the optimized finite-rank approximations of the two logical codewords from Sec.~\ref{sec:approximate_codeword_preparation} to define the
encoding map, and compute the optimal channel fidelity $F^{\mathrm{opt}}$ under photon loss, dephasing, and combined loss-dephasing noise. The recovery map is optimized by the semidefinite program (SDP) mentioned in Sec.~\ref{sec:noise_recovery}, with
the encoding and physical noise channel fixed. We report the corresponding optimal channel infidelity $1-F^{\mathrm{opt}}$ as a function of stellar rank $k$. Our numerical implementation is adapted from the public code provided in Ref.~\cite{arnelg_github2019}, with modifications to incorporate the finite-stellar-rank cat and GKP states considered here.

To make this comparison operational, we also include a break-even benchmark, defined as the infidelity of the corresponding unencoded qubit in the same physical noise model.
Codewords below this threshold outperform the unencoded benchmark, whereas codewords above it do not provide a net error-correction advantage.
In Fig.~\ref{fig:gate_average_infidelity_cat_gkp}, the break-even threshold is indicated by the horizontal dashed lines, and the shaded regions indicate the non-break-even regime.

Fig.~\ref{fig:gate_average_infidelity_cat_gkp} compares two representative code families: $N=2$ cat codes with different coherent-state amplitudes $\alpha$, and approximate GKP codes with various damping parameters $\Delta$.
Results for cat codes with higher rotation orders $(N=3,4)$ are provided in Appendix~\ref{appendix:performance of codewords}.
In general, increasing the stellar rank improves the approximation of the target codewords and tends to lower optimal channel infidelity, although the improvement is not strictly monotonic.

The insets highlight the stellar-rank intervals where the break-even crossings occur. For the $N=2$ cat codes (upper row), insets are shown for all three noise channels. Under photon loss with $\kappa t=10^{-2}$,  the $\alpha = 1.537$ curve crosses below break-even already at low stellar rank, at $k=4$, while  the $\alpha = 2.345$, and $\alpha = 2.94$ curves  cross at higher stellar rank, at $k=7$ and $k=10$, respectively. Under dephasing with $\kappa_{\phi}t=10^{-2}$, the $\alpha=1.537$ cat does not cross break-even within the range shown, reflecting its weaker phase-space separation and hence weaker protection against  dephasing. The $\alpha = 2.345$ curve crosses at intermediate stellar rank, at $k=7$, while the $\alpha = 2.94$ curve crosses at $k=10$. Under combined loss-dephasing noise with
$\kappa t=\kappa_{\phi}t=10^{-2}$, all three approximate cat codes cross break-even at same stellar rank as in the pure loss channel.

For the GKP codes (lower row), an inset is shown only for the photon-loss channel. At $\kappa t=10^{-2}$, all
three finite-rank GKP approximations already cross below the break-even threshold at relatively small stellar rank: the $\Delta=0.3$, $0.25$, and
$0.2$ curves first cross at $k=3$, $k=4$, and $k=4$, respectively. Thus, finite-rank GKP approximation can already provide an operational error-correction advantage under photon loss even though  their stellar fidelities are still far from unity. By contrast, for the dephasing and combined-noise channels, the finite-rank GKP approximations do not surpass the break-even with the range of stellar ranks considered here. Thus, in this parameter regime, the break-even advantage
of finite-rank GKP approximations is clearly observed for photon loss, but not for dephasing-dominated noise.

Beyond the break-even crossing, Fig. \ref{fig:gate_average_infidelity_cat_gkp} also shows small non-monotonic features in the channel infidelity as a function of stellar rank, most visibly for the GKP approximations under photon loss. This reflects the fact that the state-preparation objective and the QEC performance criterion are not identical. The approximate codewords are optimized to maximize the stellar fidelity of each target state separately, whereas QEC performance depends on the encoded subspace formed by the pair of logical states and on its recovery after noise. Consequently, increasing the individual stellar fidelities does not necessarily produce a monotonic improvement in logical error correction.

A related but distinct manifestation of this finite-resource effect appears in the relative ordering of different GKP approximations under photon loss. Without a stellar-rank constraint, smaller $\Delta$ gives a better finite-energy GKP code: for the pure-loss channel at $\kappa t=10^{-2}$, the optimal channel infidelities of target codes are $2.96\times 10^{-8}$, $1.08\times 10^{-7}$, and $2.23\times 10^{-6}$ for $\Delta=0.2$, $0.25$, and $0.3$, respectively, consistent with previous optimal-recovery benchmarks for finite-energy GKP codes~\cite{Albert2018PRA}.

Once a finite stellar-rank constraint is imposed, however, this ordering is reversed over much of the photon-loss panel in Fig.~\ref{fig:gate_average_infidelity_cat_gkp}: the more strongly damped $\Delta=0.3$ approximations can achieve lower infidelity than the less
damped $\Delta=0.2$ approximations at the same stellar rank. This reversal is caused by finite-rank preparation error.  Indeed, up to $k=40$, the optimized
stellar fidelities for $\Delta=0.2$ reach only $0.981$ for
$|\mathrm{GKP}^{0}\rangle$ and $0.973$ for $|\mathrm{GKP}^{1}\rangle$, whereas for
$\Delta=0.3$ both logical codewords reach nearly unit stellar fidelity, as shown in Fig. \ref{fig:stellar_profile_wignerfunction_gkp}.

\section{QEC with optimized encodings at finite stellar rank}
\label{sec:QEC_optimal_codewords}

The results of the previous section were obtained by approximating fixed target codewords under stellar-rank constraints.
However, stellar fidelity quantifies the quality of individual state preparation, not the properties of the encoded subspace after noise and recovery, so high stellar fidelity need not translate into high logical performance.
This motivates directly optimizing the encoding itself under a stellar-rank constraint, rather than fixing the codewords a priori.
Similar ideas have been explored in Ref.~\cite{Leviant2022quantum}, where codewords were optimized under a fixed average energy constraint; here we instead impose a fixed stellar rank.

\subsection{General considerations on stellar-rank-constrained encoding optimization}
Any single-mode pure state $\ket{\psi}$ with finite stellar rank can be decomposed as a Gaussian unitary acting on a unique core state $\ket{C_\psi}$ with the same stellar rank \cite{ulysse2020PRL},
\begin{equation}
    \ket{\psi} = \hat{G}(\zeta,\beta)\ket{C_\psi},
    \label{eq:core_states}
\end{equation}
where $\hat{G}(\zeta,\beta)$ is a Gaussian unitary.
For finite $r^\star(\psi)$, the core state has finite Fock support,
\begin{equation}
    \ket{C_\psi} = \sum_{n=0}^{r^\star(\psi)} c_n \ket{n}.
\end{equation}
This decomposition provides a natural variational parameterization of states with stellar rank at most $k$: one jointly optimizes the Gaussian parameters $(\zeta,\beta)$ and the core-state coefficients
$\{c_n\}_{n=0}^{k}$ (subject to normalization), using the same Gaussian search window as in the stellar-fidelity optimization above.

For encoding, we use an isometry $V$ from the logical qubit space to the bosonic mode. We restrict the image of $V$ to be a Gaussian transform of a two-dimensional core subspace of  $\mathrm{span}\{\ket{0},\ldots,\ket{k}\}$. Thus any encoded logical superposition $V(a\ket{0_L}+b\ket{1_L})$, with $|a|^2+|b|^2=1$, has stellar rank at most $k$.

We define the stellar-rank cost of the encoding as 
\begin{equation}
    r^\star(V)=
    \max_{\ket{\psi_L}}
    r^\star\!\left(V\ket{\psi_L}\right),
\end{equation}
where the maximization is over all normalized logical pure states. The ideal constrained optimization is then 
\begin{equation}
    V^{\mathrm{opt}}=
    \underset{V\,:\,r^\star(V)\le k}{\mathrm{argmax}}\;
    F^{\mathrm{opt}}(\mathcal{N}\circ\mathcal{E}_V),
    \label{eq:encoding_opt}
\end{equation}
where $\mathcal{E}_V(\hat{\rho}_L)=V\hat{\rho}_L V^\dagger$.
Here
\begin{equation}
    F^{\mathrm{opt}}(\mathcal{N}\circ\mathcal{E}_V)
    =\max_{\mathcal{R}}
    F(\mathcal{R}\circ\mathcal{N}\circ\mathcal{E}_V)
\end{equation}
is the channel fidelity optimized over the recovery map. For fixed $V$, this recovery optimization is an SDP \cite{Kosut2009QIP}. The remaining optimization over $V$, however, is non-convex because of the finite-stellar-rank constraint and the Gaussian-core parametrization.

Solving the recovery SDP on top of the variational search is computationally costly and did not give stable convergence in the parameter regime studied here. We therefore use the Petz-recovery fidelity as a surrogate objective for optimizing the encoding. For a fixed encoding, the Petz recovery is defined analytically from the noise channel and the code space \cite{Petz1988QJM}, and satisfies the near-optimality bound \cite{Zheng2024PRL}
\begin{equation}
    \frac{1}{2}(1-\tilde{F}^{\mathrm{opt}})
    \leq
    1-F^{\mathrm{opt}}
    \leq
    1-\tilde{F}^{\mathrm{opt}},
    \label{eq:petz_bound}
\end{equation}
where $F^{\mathrm{opt}}$ is the  optimal channel fidelity achieved
by the optimal recovery, and $\tilde{F}^{\mathrm{opt}}$ is the
Petz-recovery fidelity, given by
\begin{equation}
    \tilde{F}^{\mathrm{opt}}
    =
    \frac{1}{d_L^2}
    \left\|
        \mathrm{Tr}_L \sqrt{M}
    \right\|_F^2,
    \label{eq:Petz_fidelity}
\end{equation}
where $d_L=2$ and  $M_{[\mu l],[\nu k]} = \langle \mu_L | \hat{E}_l^\dagger \hat{E}_k | \nu_L \rangle$ is the QEC matrix constructed from Kraus error operators $\hat{E}_k$ associated with the physical noise channel, and $\|\cdot\|_F$ denotes the Frobenius norm. 

For the numerical search, we optimize
\begin{equation}
    V^{\mathrm{Petz}}
    =
    \underset{V\,:\,r^\star(V)\le k}{\mathrm{argmax}}
    \;
    \tilde{F}(\mathcal{N}\circ\mathcal{E}_V).
    \label{eq:encoding_opt_petz}
\end{equation}
This remains a non-convex variational optimization. 
After the search, the resulting code states are fixed, and their final performance is evaluated with the optimal recovery SDP. 
Thus, the Petz recovery is used only to guide the encoding search, while the benchmark reported below is the optimal-recovery infidelity $1-F^{\mathrm{opt}}$.

\subsection{Numerically optimized encoding performance and phase-space structure}
\begin{figure*}[t]
    \centering    \includegraphics[width=\textwidth]{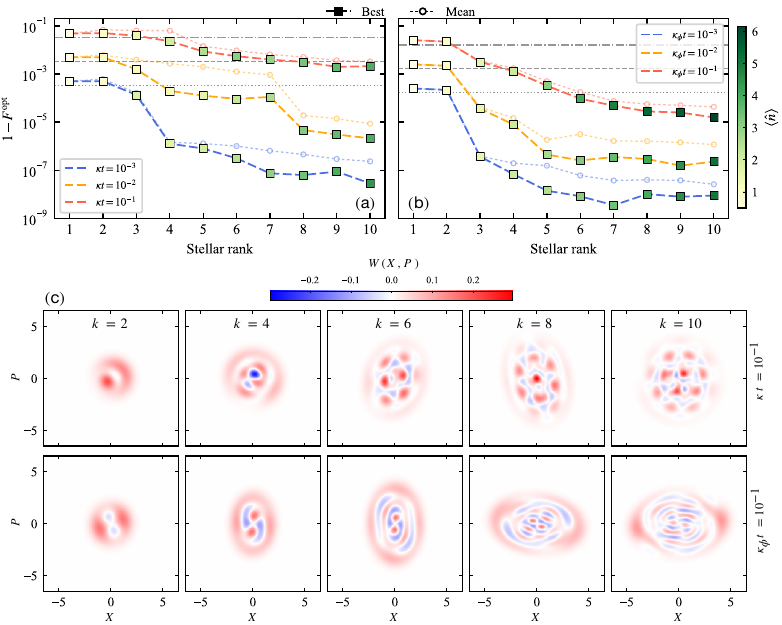}
    \caption{Optimized error-correction performance, average photon number, and phase-space structure of finite-stellar-rank bosonic encodings under photon-loss and dephasing noise.
    (a) Optimal-recovery channel infidelity $(1-F^{\mathrm{opt}})$ as a
    function of stellar rank $k$ for encodings optimized under photon loss with $\kappa t\in\{10^{-3},10^{-2},10^{-1}\}$.The Petz-recovery fidelity is used as the surrogate objective during the encoding search, while the plotted infidelities are computed using optimal recovery for the final codewords.
    Square markers show the best result over independent numerical runs, while
    open circles show the mean over runs. Marker colors indicate the mean photon
    number $\langle\hat{n}\rangle$ of the corresponding optimized encodings.
    Gray horizontal lines mark the break-even infidelity for the three noise strength, with dotted, dashed, and dashed-dotted line corresponding to $10^{-3},10^{-2},10^{-1}$, respectively.
    (b) Same as (a) for dephasing noise with
    $\kappa_{\phi}t\in\{10^{-3},10^{-2},10^{-1}\}$.
     (c) Wigner functions of the maximally mixed encoded state for $k=2,4,6,8,10$ at error rate $10^{-1}$. The upper row corresponds to photon loss with $\kappa t=10^{-1}$, and the lower row to dephasing with $\kappa_{\phi}t=10^{-1}$.}
    \label{fig:entanglement_fid_optimal_states}
\end{figure*}

For each noise channel and noise strength, we solve the optimization problem in Eq.~\eqref{eq:encoding_opt_petz}
over a range of stellar ranks. The two logical core states are represented by a
$(k+1)\times 2$ coefficient matrix and orthonormalized by QR decomposition \cite{Golub2013Matrix} before applying a Gaussian unitary. All states and noise channels are represented in a truncated Fock space with
$d_{\mathrm{phys}}=4k+2$ for $k\geq 2$ and $d_{\mathrm{phys}}=10$ for $k=1$. The cutoff is chosen to keep the optimization computationally tractable while leaving sufficient Hilbert-space support for the Gaussian-transformed finite-rank core states. 

Photon loss is implemented with truncated Kraus error operators up to loss order
$K_{\mathrm{loss}}=30$, while dephasing is implemented using the exact
finite-dimensional spectral Kraus representation of the photon-number dephasing
kernel. The Kraus operators are renormalized to satisfy trace preservation
within the truncated Hilbert space.

The optimization is performed using Adaptive Moment Estimation (Adam) algorithm with learning rates
$3\times 10^{-4}$ for both the core and Gaussian parameters, for $4000$
optimization steps. To reduce trapping in local optima, each value of $k$ is
optimized from $S=60$ independent random initializations. A weak photon-number
penalty and a tail penalty are included to suppress solutions near the Fock
cutoff. As a cutoff check, when available we select the best solution whose total population in
the highest six Fock levels is below $10^{-5}$.

For each run $j$, the encoding is first obtained by maximizing the Petz-recovery fidelity. We then compute the optimal-recovery channel fidelity $F^{\mathrm{opt}}_j$ for the final encodings.
The results are summarized by the best value
\begin{equation}
    F^{\mathrm{best}}=\max_{1\leq j\leq S} F^{\mathrm{opt}}_{j},
\end{equation}
and the mean value
\begin{equation}
     F^{\mathrm{avg}}=\frac{1}{S}\sum_{j=1}^{S} F^{\mathrm{opt}}_{j}.
\end{equation}
In Figs. 4(a) and (b), we plot the corresponding optimal-recovery infidelities: 1-$F^{\mathrm{best}}$, shown with filled square markers connected by long dashed lines, and 1-$F^{\mathrm{avg}}$, shown with open circle markers connected by short dashed lines.

For both noise channels, increasing the stellar rank generally improves the optimized performance, reflecting the larger non-Gaussian variational space available for constructing noise-adapted bosonic codewords. Small oscillations appear in the best-run curves, especially at low physical error rates and larger stellar ranks, where the infidelity reaches the $10^{-8}$--$10^{-9}$ level, and differences between nearby local optima become visible on a  logarithmic scale. The mean over random initializations is therefore included to provide a clearer view of the overall trend with stellar rank.

The break-even behavior depends strongly on the noise model. Under dephasing [Fig.~\ref{fig:entanglement_fid_optimal_states}(b)], optimized encodings surpass the break-even benchmark already at $k=2$ for the noise strengths considered. Under photon loss [Fig.~\ref{fig:entanglement_fid_optimal_states}(a)],
the required stellar rank increases with the loss rate. For weak loss, the break-even threshold is crossed at low stellar rank, whereas for
$\kappa t=10^{-1}$ larger stellar rank is required. This reflects the competing requirements of loss protection: the encoded states must develop enough phase-space structure to encode information robustly, while avoiding excessive photon occupation that would increase sensitivity to excitation loss.

To compare the codes at a common resource scale, we focus on $k=4$ and $\kappa t=10^{-2}$. For the smallest-amplitude $N=2$ cat code in
Fig.~\ref{fig:gate_average_infidelity_cat_gkp}, with $\alpha=1.537$, the
finite-stellar-rank approximation first crosses break-even at $k=4$, with
$1-F^{\mathrm{opt}}=3.41\times 10^{-4}$.
At the same stellar-rank scale, the optimized encoding found in this section
reaches $1-F^{\mathrm{opt}}=1.99\times 10^{-4}$.

We also compare with the corresponding  binomial code~\cite{Michael2016PRX} with same stellar rank,
\begin{equation}
    \ket{0_L^{\mathrm{bin}}}=
    \frac{1}{\sqrt{2}}\left(\ket{0}+\ket{4}\right),
    \qquad
    \ket{1_L^{\mathrm{bin}}}=\ket{2}.
    \label{eq:N2_binomial_code}
\end{equation}
Under the same optimal-recovery benchmark, this code gives
$1-F^{\mathrm{opt}}=2.9\times 10^{-3}$ \cite{Albert2018PRA}.
Thus, in this representative comparison, the optimized finite-rank encoding achieves the lowest infidelity among the three codes, suggesting that
noise-adapted optimization can improve performance within a fixed stellar-rank budget.
The Wigner functions of the corresponding maximally mixed encoded states are shown in Appendix~\ref{appendix:optimal_states}, providing a
basis-independent comparison of their phase-space structures.

The mean photon number $\langle\hat{n}\rangle$, indicated by the marker color, further illustrates the different energetic strategies favored by the two noise channels. Under photon loss, stronger noise favors lower-energy encodings, since states with larger photon occupation are more susceptible to excitation decay; nevertheless, at fixed loss rate the average photon number can still increase with stellar rank as additional non-Gaussian resources enable more structured codewords.
Under dephasing, the optimal photon number increases with the dephasing rate, as encoded states with broader photon-number support and larger phase-space separation become favorable, similar to the protection mechanism underlying cat codes \cite{Arne2020PRX}.

These trends are directly reflected in the structure of the optimized code spaces.
Fig.~\ref{fig:entanglement_fid_optimal_states}(c) shows the Wigner functions of the logical maximally mixed state
\begin{equation}
\hat{\rho}_{\mathrm{mix}}
=
\frac{1}{2}
\left(
\ketbra{0_L}{0_L}
+
\ketbra{1_L}{1_L}
\right),
\end{equation}
at loss error rate $\kappa t  = 10^{-1}$ and dephasing error rate $\kappa_\phi t  = 10^{-1}$ for representative stellar ranks.
The full set of optimized states across different error rates is provided in Appendix~\ref{appendix:optimal_states}.

Under photon loss (upper row), the optimized states evolve from localized low-rank structures toward increasingly grid-like Wigner functions at larger $k$, resembling approximate GKP-type states~\cite{Timo2022PRX.Q, Leviant2022quantum}.
By contrast, under dephasing noise (lower row), the optimized states develop approximate $\pi$-rotation symmetry, consistent with the broader phase-space separation favored by dephasing protection~\cite{Arne2020PRX}.

Overall, these results show that stellar rank controls both the achievable logical performance and the geometry of the optimized bosonic codewords. Increasing the available non-Gaussian resource budget enables more structured, noise-adapted encodings, but the optimal structure depends strongly on the dominant noise mechanism. Photon loss favors a balance between phase-space structure and low energy, whereas dephasing favors broader, more separated codewords with stronger rotational structure.

\section{Discussion and conclusions}
\label{sec:conclusion}
In this work, we developed a resource-constrained framework for bosonic QEC under finite non-Gaussian resources, using stellar rank, the minimal number of photon additions required to generate a target state, as the key resource measure. 

For fixed code families, we showed that the difficulty of finite-rank approximation is controlled by the phase-space complexity of the target states: larger-amplitude cat states and less damped GKP states require higher stellar rank to approximate accurately.
This gives rise to a finite-resource trade-off in which codewords that are optimal in the ideal limit can become suboptimal when stellar rank is limited, as preparation error outweighs intrinsic code quality.
Nevertheless, finite-rank approximations can already surpass break-even before exact preparation is achievable, with the  required stellar rank depending on both the code family and the noise channel.

Going beyond fixed targets, we directly optimized code spaces within finite-stellar-rank variational classes. For computational tractability, the search was guided by the Petz-recovery fidelity, while the final encodings were benchmarked using optimal recovery. The optimized encodings reveal noise-adapted code structures: grid-like encodings emerge under photon loss, while approximately rotation-symmetric structures arise under dephasing, consistent with previous studies based on codewords optimization at fixed average photon number \cite{Leviant2022quantum}.
The optimized average photon number also adapts to the dominant noise mechanism, balancing phase-space structure against energy cost under photon loss and favoring broader photon-number support under dephasing.
In particular, for the optimized encodings, stellar rank $k=2$ is sufficient to surpass break-even for all dephasing strengths considered, whereas under photon loss the required stellar rank increases with the loss rate.

Taken together, our results establish stellar rank as an operationally meaningful and experimentally motivated resource measure for bosonic QEC, bridging non-Gaussian resource theory and logical error-correction performance under realistic state-preparation constraints. 
A natural next step is to develop more systematic optimization methods for finite-stellar-rank code spaces and to compare stellar rank with other non-Gaussianity resource measures \cite{takagi2018PRA, albarelli2018PRA, hahn2025quantum, dias2024PRA, calcluth2025classical}, clarifying how different non-Gaussian constraints affect
state preparation and logical performance in bosonic QEC.

Just before completion of our work, we became aware of a related work dealing with approximations of cat states with finite stellar rank \cite{Julian2026arxiv}
\section{Acknowledgements}
We thank Alex Maltesson, Rivu Gupta and Francesco Arzani for useful discussions. 
 G.F., U.C.\ and A.F.\ acknowledge funding from the European Union’s Horizon Europe Framework Programme (EIC Pathfinder Challenge project Veriqub) under Grant Agreement No.\ 101114899.
G.F.\ acknowledges financial support from the Swedish Research Council (Vetenskapsradet) through the project grant DAIQUIRI, as well as from the Olle Engkvist foundation. G.F., R.W.\ and T.H.\ acknowledge support from the Knut and Alice Wallenberg Foundation through the Wallenberg Center for Quantum Technology (WACQT). R.W. acknowledges resources at the Chalmers Centre for Computational Science and Engineering (C3SE).
T.H.\ acknowledges support from Defence Science and Technologies Group (DSTG) and Advanced Strategic Capabilities Accelerator (ASCA) through its Emerging and Disruptive Technologies (EDT) Program. The authors used OpenAI ChatGPT (GPT-5.5) as an auxiliary tool for improving the clarity and readability of the manuscript and checking code readability. All AI-assisted output was reviewed and verified by the authors, who take full responsibility for the content of this work.

\appendix
\section{QEC framework and noise channels}

In this Appendix, we summarize the Kraus-operator formulation of the bosonic quantum error correction protocol considered in the main text, together with the explicit noise channels used in our simulations.
\subsection{General QEC framework}
\label{appendix:qec-framework}
A bosonic QEC protocol consists of three stages: encoding, physical noise, and recovery. Logical information is initially defined on a finite-dimensional logical Hilbert space $\mathcal H_S$ and encoded into the bosonic Hilbert space $\mathcal H_C$.

The encoding, noise, and recovery processes are described by completely positive trace-preserving (CPTP) maps~\cite{Kosut2009QIP},
\begin{align}
    \mathcal{E}(\hat{\rho}_S)
    &= \sum_s \hat{S}_s \hat{\rho}_S \hat{S}_s^\dagger, \\
    \mathcal{N}(\hat{\rho}_C)
    &= \sum_e \hat{E}_e \hat{\rho}_C \hat{E}_e^\dagger, \\
    \mathcal{R}(\hat{\sigma}_C)
    &= \sum_r \hat{R}_r \hat{\sigma}_C \hat{R}_r^\dagger,
\end{align}
where $\hat{\rho}_S\in\mathcal H_S$ denotes a logical input state and $\hat{\sigma}_C\in\mathcal H_C$ denotes a bosonic state after the action of the noise channel.

The Kraus operators satisfy the trace-preserving conditions
\begin{equation}
    \sum_s \hat{S}_s^\dagger \hat{S}_s = I_S,
    \qquad
    \sum_e \hat{E}_e^\dagger \hat{E}_e = I_C,
    \qquad
    \sum_r \hat{R}_r^\dagger \hat{R}_r = I_C,
\end{equation}
where $I_S$ and $I_C$ are the identity operators on the logical and bosonic Hilbert spaces, respectively.

The resulting effective logical channel is given by
\begin{equation}
    \mathcal{C}
    =
    \mathcal{R}\circ\mathcal{N}\circ\mathcal{E}.
\end{equation}
Acting on an input logical state $\hat{\rho}_S$, the channel takes the explicit form
\begin{equation}
    \mathcal{C}(\hat{\rho}_S)
    =
    \sum_{r,e,s}
    \hat{C}_{r,e,s}
    \hat{\rho}_S
    \hat{C}_{r,e,s}^\dagger .
\end{equation}
where
\begin{equation}
    \hat{C}_{r,e,s}
    =
    \hat{R}_r \hat{E}_e \hat{S}_s .
\end{equation}

For a given encoding and noise channel, the recovery map can be optimized to maximize the channel fidelity discussed in the main text.
In Sec.~\ref{sec:QEC_optimal_codewords}, we instead employ the Petz recovery map as a computationally efficient near-optimal recovery channel, which enables direct optimization of finite-stellar-rank bosonic codewords.

We now summarize the explicit noise models considered in this work.

\subsection{Photon-loss channel}
\label{appendix:photon-loss}
The photon-loss channel for a single bosonic mode with annihilation operator $\hat a$ is generated by the Lindblad master equation
\begin{equation}
    \frac{d\hat{\rho}}{dt}
    =
    \kappa\,\mathcal{D}[\hat a](\hat{\rho}),
\end{equation}
where
\begin{equation}
    \mathcal{D}[\hat L](\hat{\rho})
    =
    \hat L \hat{\rho} \hat L^\dagger
    -\frac{1}{2}\hat L^\dagger \hat L \hat{\rho}
    -\frac{1}{2}\hat{\rho}\hat L^\dagger \hat L .
\end{equation}

The corresponding CPTP map admits the Kraus decomposition~\cite{Grimsmo2021PRX.Q}
\begin{equation}
    \mathcal{N}_{\mathrm{loss}}(\hat{\rho})
    =
    \sum_{\ell=0}^{\infty}
    \hat A_\ell
    \hat{\rho}
    \hat A_\ell^\dagger ,
\end{equation}
with Kraus operators
\begin{equation}
    \hat A_\ell
    =
    \frac{(1-e^{-\kappa t})^{\ell/2}}{\sqrt{\ell!}}
    e^{-\kappa t \hat n /2}
    \hat a^\ell .
\end{equation}

\subsection{Dephasing channel}
\label{appendix:dephasing}
The dephasing channel is generated by the number operator $\hat n=\hat a^\dagger \hat a$,
\begin{equation}
    \frac{d\hat{\rho}}{dt}
    =
    \kappa_\phi\,\mathcal D[\hat n](\hat{\rho}).
\end{equation}

Its Kraus decomposition is~\cite{Grimsmo2021PRX.Q}
\begin{equation}
    \mathcal{N}_{\mathrm{deph}}(\hat{\rho})
    =
    \sum_{l=0}^{\infty}
    \hat B_l
    \hat{\rho}
    \hat B_l^\dagger ,
\end{equation}
where
\begin{equation}
    \hat B_l
    =
    \frac{(\kappa_\phi t)^{l/2}}{\sqrt{l!}}
    e^{-\kappa_\phi t \hat n^2/2}
    \hat n^l .
\end{equation}

Equivalently, in the Fock basis,
\begin{equation}
    \mathcal{N}_{\mathrm{deph}}(\ketbra{m}{n})
    =
    e^{-\frac{\kappa_\phi t}{2}(m-n)^2}
    \ketbra{m}{n}.
\end{equation}

\section{Analytical derivation of stellar fidelity for cat and GKP codes}
\label{appendix:deviation_stellar_fidelity}

In this Appendix, we provide analytical derivations of the stellar fidelity for cat states and GKP states.
We refer to the definition of stellar fidelity and the associated Gaussian optimization introduced in Eq.~ (\ref{eq:def_stellar_fidelity}) of the main text.

\subsection{Cat codes}
\label{appendix:stellar_fidelity_cat}
For the cat codes, we derive an analytically simplified expression for the stellar fidelity of them, that will be used for the numerical evaluation.
First, we consider the codewords of cat codes given in Eq.  (\ref{eq:def_general_cat_codes}).
With Eq.~(\ref{eq:stellar_fidelity_cat}) and 
 the squeezing parameter $\zeta=re^{i\theta}$ and displacement $\beta\in\mathbb{C}$, one has
\begin{align}
    \bra{m}&\hat{S}(\zeta)\hat{D}(\beta)\ket{\tilde{\alpha}}\nonumber\\ 
    &= \frac{1}{\sqrt{m!}} 
       e^{\tfrac{1}{2}(\tilde{\alpha}^\star\beta-\tilde{\alpha}\beta^\star)} \,
       \bra{0}\hat{a}^m \hat{S}(\zeta)\hat{D}(\beta+\tilde{\alpha})\ket{0}.
\end{align}
Here we use the displacement relation
\begin{align}
\hat{D}(\beta)|\alpha\rangle
=
e^{\frac12(\alpha\beta^\star-\alpha^\star\beta)}
|\alpha+\beta\rangle.
\label{eq:displacement_relation}
\end{align}
Then we switch to the stellar representation \cite{ulysse2020PRL},
\begin{align}
    \bra{m}\hat{S}(\zeta)\hat{D}(\beta)\ket{\tilde{\alpha}} 
    &= \frac{1}{\sqrt{m!}}
       \Big[ \partial^m_y \,
       e^{\tfrac{1}{2}(\tilde{\alpha}^\star\beta-\tilde{\alpha}\beta^\star)}
       G^\star_{\zeta, \tilde{\alpha}+\beta}(y)\Big]_{y=0},
\end{align}
with
\begin{align}
    G^\star_{\zeta, \tilde{\alpha}+\beta}(y) 
    &= (1-|a|^2)^{1/4} 
    \exp\!\Big(-\tfrac{1}{2}ay^2 + by + c \Big), \nonumber\\
    a &= e^{i\theta}\tanh r, \quad 
    b = \tfrac{\tilde{\alpha}+\beta}{\cosh r}, \nonumber\\
    c &= \tfrac{1}{2}e^{i\theta}\tanh r (\tilde{\alpha}+\beta)^2
         - \tfrac{1}{2}|\tilde{\alpha}+\beta|^2. 
\label{eq:gaussian_stellar_representation}
\end{align}
Using the identity
\begin{equation}
    \Big[\partial^m_y e^{-\tfrac{1}{2}ay^2+by}\Big]_{y=0} 
    = a^{m/2}{\rm {H}e}_m\!\left(\tfrac{b}{\sqrt{a}}\right),
    \label{eq:Hemitian_polynomial}
\end{equation}
where ${\rm {H}e}_m,$ is the $m$-th Hermite polynomial, we finally obtain
\begin{align}
     \bra{m}\hat{S}(\zeta)\hat{D}(\beta)\ket{\tilde{\alpha}}
     &= \frac{1}{\sqrt{c_r m!}}
     e^{-\tfrac{1}{2}(|\tilde{\alpha}|^2+|\beta|^2)-\tilde{\alpha}\beta^\star
     +\tfrac{1}{2}e^{i\theta}t_r(\tilde{\alpha}+\beta)^2} \nonumber \\
     &\qquad \times (e^{-i\theta}t_r)^{m/2}
     {\rm {H}e}_m\!\Big(\tfrac{\tilde{\alpha}+\beta}{\sqrt{s_rc_r}}e^{i\theta/2}\Big),
\end{align}
where $c_r=\cosh r$, $s_r=\sinh r$, $t_r=\tanh r$.  

Thus, the stellar fidelities for the logical   cat code states $|\mathrm{Cat}_{N,\alpha}^\mu\rangle$ are
\begin{align}
    &f^\star_k(|\mathrm{Cat}_{N,\alpha}^{\mu}\rangle)\nonumber\\ 
    &= \max_{\hat{G}\in \mathcal{G}}
       \frac{e^{-(|\alpha|^2+|\beta|^2)}}{c_r\mathcal{N}_{\mu}}
       \sum_{m=0}^k \frac{t_r^m}{m!}
       \Big|\sum_{z=0}^{2N-1} (-1)^{\mu z}u_m(\tilde{\alpha},\zeta,\beta)\Big|^2 ,
\end{align}
where
\begin{equation}
   u_m(\tilde{\alpha}, \zeta, \beta) =  
   e^{-\tilde{\alpha}\beta^\star+\tfrac{1}{2}e^{i\theta}t_r(\tilde{\alpha}+\beta)^2}
   {\rm {H}e}_m\!\left(\tfrac{\tilde{\alpha}+\beta}{\sqrt{s_rc_r}}e^{i\theta/2}\right).
\end{equation}
This expression makes it possible to evaluate analytically the stellar fidelity of a target cat codeword for arbitrary rotational symmetry $N$.

\subsection{GKP codes}
\label{appendix:stellar_fidelity_GKP}
The approximate GKP logical states are defined in the main text see Eq.~(\ref{eq:codewords-GKP}), where they are expressed in the coherent-state basis with a Gaussian damping parameter $\Delta$.
Then the stellar fidelity of the approximate logical codewords states can be obtained by the definition of stellar fidelity Eq.~(\ref{eq:def_stellar_fidelity}),
\begin{align}
f^\star_k(\ket{\text{GKP}^\mu_\Delta})
=
\max_G
\sum_{m=0}^{k}
\lvert
\langle m | \hat{G} \ket{\text{GKP}^\mu_\Delta}
\rvert^2.
\end{align}
and the relevant matrix elements become
\begin{align}
\langle m|\hat{G}\ket{\text{GKP}^\mu_\Delta}
&=\frac{1}{\sqrt{\mathcal{N}_\mu}}
\sum_{g,l\in \mathbb{Z}}
e^{-i\pi l (g + \frac{\mu}{2})}
e^{-\Delta^2 |\alpha_{gl}|^2}\nonumber\\
& \times \langle m|\hat{G}
|e^{-\Delta^2}\alpha_{gl}\rangle.
\end{align}
We decompose the Gaussian unitary $\hat{G}$ as the displacement  $\hat{D}(\zeta)$ and squeezing operators $\hat{S}(\xi)$, respectively.
Then with the displacement relation Eq.~(\ref{eq:displacement_relation}),
one obtains
\begin{align}
&\langle m|\hat{G}\ket{\text{GKP}^\mu_\Delta}\nonumber\\
&=
\frac{1}{\sqrt{\mathcal{N}_\mu}}
\sum_{g,l \in \mathbb{Z}}
e^{-i\pi l (g + \frac{\mu}{2})}
e^{-\Delta^2 |\alpha_{gl}|^2}
e^{\frac12(\tilde{\alpha}_{gl}\zeta^\star
-\tilde{\alpha}_{gl}^\star\zeta)} \nonumber\\
&\times\langle m|
\hat{S}(\xi)\hat{D}(\tilde{\alpha}_{gl}+\zeta)
|0\rangle,\nonumber\\
& = \frac{1}{\sqrt{\mathcal N_\mu\cdot m!}}\sum_{g,l\in \mathbb{Z}}
e^{-i\pi l (g + \frac{\mu}{2})}
e^{-\Delta^2|\alpha_{gl}|^2}
e^{\frac12( \tilde{\alpha}_{gl}^\star\zeta-\tilde{\alpha}_{gl}\zeta^\star)}\nonumber\\
&\times \langle 0|\hat{a}^m
\hat{S}(\xi)\hat{D}(\tilde{\alpha}_{gl}+\zeta)|0\rangle,
\end{align}
where $\tilde{\alpha}_{gl}=e^{-\Delta^2}\alpha_{gl}$ and we used that $\langle m|\hat{S}(\xi)\hat{D}(\tilde{\alpha}_{gl} + \zeta)|0\rangle
=
\frac{1}{\sqrt{m!}}
\langle 0|
\hat{a}^m
\hat{S}(\xi)
\hat{D}(\tilde{\alpha}_{gl} + \zeta)
|0\rangle.$
Similarly, switching to the stellar representation \cite{ulysse2020PRL}, we obtain
\begin{align}
&\bra{m}\hat{G}\ket{\text{GKP}^\mu_\Delta} \nonumber\\
&=
\frac{1}{\sqrt{ \mathcal N_\mu \cdot m!}}
\sum_{g,l\in \mathbb{Z}}
e^{-i\pi l (g + \frac{\mu}{2})}
e^{-\Delta^2|\alpha_{gl}|^2} e^{\frac12( \tilde{\alpha}_{gl}^\star\zeta-\tilde{\alpha}_{gl}\zeta^\star)}\nonumber\\
& \times 
\left[
\partial_z^m
G^\star_{\xi,\tilde{\alpha}_{gl} + \zeta}(z)
\right]_{z=0}.   
\label{eq:matrix_element_GKP}
\end{align}

With Eq.~\eqref{eq:gaussian_stellar_representation}  and Eq.~\eqref{eq:Hemitian_polynomial} we obtain 
\begin{align}
&\bra{m}\hat{G}\ket{\text{GKP}^\mu_\Delta}\nonumber\\
    &=
    \frac{1}{\sqrt{\mathcal N_\mu \cdot c_r \cdot m!}}
    \sum_{g,l\in \mathbb{Z}}
    e^{-i\pi l (g + \frac{\mu}{2})}
    e^{-\Delta^2|\alpha_{gl}|^2} e^{-\frac12(|\tilde{\alpha}_{gl}|^2+|\zeta|^2)}
    \nonumber\\
    &\times e^{
    -\tilde{\alpha}_{gl} \zeta^\star
    +\tfrac12 t_r\, e^{i\theta}
    (\zeta+\tilde{\alpha}_{gl})^2}\left(e^{i\theta}t_r\right)^{m/2}{\rm {H}e}_m\!\left(
    \frac{\zeta+\tilde{\alpha}_{gl}}
    {\sqrt{c_rs_r}}\,
    e^{i\theta/2}
    \right).
\end{align}
Finally, we have the general stellar fidelity of GKP logical states:
\begin{align}
&f^\star_k(\ket{\text{GKP}^\mu_\Delta})\nonumber\\
    &=
    \max_{\hat G}
    \frac{1}{\mathcal N_\mu \cdot c_r}
    \sum_{m=0}^{k}
    \frac{t_r^m}{m!}
    \Big|
    \sum_{g,l\in \mathbb{Z}}
    e^{-i\pi l (g + \frac{\mu}{2})}
    e^{-\Delta^2|\alpha_{gl}|^2}
    \nonumber\\
    &
    \times e^{-\frac12(|\tilde{\alpha}_{gl}|^2+|\zeta|^2)-\tilde{\alpha_{gl}} \zeta^\star}
    e^{\frac12 e^{i\theta}t_r(\zeta+\tilde{\alpha}_{gl})^2}\nonumber\\
    & \times {\rm {H}e}_m\!\left(
    \frac{\zeta+\tilde{\alpha}_{gl}}
    {\sqrt{c_r\,s_r}}
    e^{i\theta/2}
    \right)
    \Big|^2.
\end{align}

\section{Stellar fidelity plateaus of cat codes}
\label{appendix:fidelity_platuea_cat}

In this Appendix, we analyze the origin of the plateau structures observed in the stellar-fidelity profiles of cat states.
Consider the logical cat state in Eq.~\eqref{eq:def_general_cat_codes}.
In the large-amplitude regime, the coherent components become well separated in phase space and their mutual overlaps are exponentially suppressed.
The state can then be approximated as a superposition of $2N$ nearly orthogonal coherent components with equal weights,
\begin{equation}
    |\mathrm{Cat}_{N,\alpha}^{\mu}\rangle
    \approx
    \frac{1}{\sqrt{2N}}
    \sum_{z=0}^{2N-1}
    (-1)^{\mu z}
    \ket{\alpha_z}, \quad (\mu = 0, 1) 
\end{equation}
where $\alpha_z= e^{iz\pi/N}\alpha$.
For a fixed stellar rank $k$, let $\hat G_o^k$ denote the Gaussian unitary that maximizes the stellar fidelity with that state.
Substituting the above expansion into  Eq.~(\ref{eq:def_stellar_fidelity}) gives
\begin{align}
    f_k^\star
    &=
    \frac{1}{2N}
    \sum_{z,z'=0}^{2N-1}
    (-1)^{\mu(z-z')}
    \bra{\alpha_{z'}}
    (\hat G_o^k)^\dagger
    \hat{\Pi}_k
    \hat G_o^k
    \ket{\alpha_z}.
\end{align}
where $\hat{\Pi}_k=\sum_{m=0}^{k}\ketbra{m}$.
\begin{figure}[t]
    \centering    \includegraphics[width=\columnwidth]{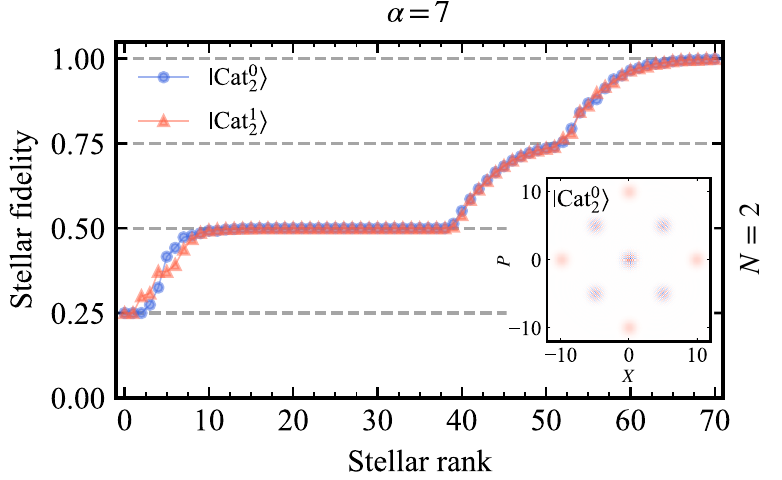}
    \caption{
    Stellar fidelity as a function of stellar rank $k$ for large-amplitude cat states with $\alpha=7$ and $N=2$. 
    The blue circles and red triangles correspond to the two logical codewords $|\rm Cat_2^0\rangle$ and $|\rm Cat_2^1\rangle$, respectively. 
    In the large-amplitude regime, the coherent components are well separated in phase space, leading to visible plateau structures in the stellar-fidelity profile. 
    For $N=2$, the expected plateau values are $1/4$, $1/2$, $3/4$, and $1$, corresponding to the effective capture of one, two, three, and four coherent components within the finite-rank approximation. The gray dotted lines indicate the plateau values.
    And the inset shows the phase-space locations of the coherent components for the one of logical cat states $|{\rm Cat_2^0}\rangle$.}
    \label{fig:stellar_fidelity_cat_large_alpha}
\end{figure}
In the large-amplitude regime, the Gaussian-transformed coherent branches remain well separated in phase space, so the off-diagonal matrix elements between distinct branches are negligible,
\begin{equation}
\bra{\alpha_{z'}}(\hat{G}_o^k)^\dagger\hat{\Pi}_k\hat{G}_o^k\ket{\alpha_z} \approx 0, \qquad z\neq z'.
\end{equation}
Retaining only the diagonal contributions, the stellar fidelity reduces to
\begin{equation}
    f_k^\star
    \approx
    \frac{1}{2N}
    \sum_{z=0}^{2N-1}
    p_z(k;\hat{G}_o^k),
\end{equation}
where
\begin{equation}
    p_z(k;\hat{G}_o^k)
    =
    \sum_{m=0}^{k}
    |\langle m|\hat{G}_o^k|\alpha_z\rangle|^2
\end{equation}
is the cumulative probability that the $z$-th Gaussian-transformed coherent component lies within the truncated Fock subspace spanned by $\{\ket{0}, ...,\ket{k}\}$.

The plateau structure arises because different coherent branches become captured at different values of $k$.
Over intervals of stellar rank where $q$ out of $2N$ branches are fully captured while the remaining $2N-q$ are not yet resolved, one has
\begin{equation}
    p_z(k;\hat{G}_o^k) \approx \begin{cases} 
    1, & z \in S_q, \\ 0, 
    & z \notin S_q, 
    \end{cases}
\end{equation}
where $S_q$ denotes the subset of $q$ captured branches.
The stellar fidelity then takes the approximate constant value
\begin{equation}
    f_k^\star \approx \frac{q}{2N}, \qquad q = 1,2,\ldots,2N,
\end{equation}
which is approximately independent of $k$ within that interval, giving rise to the observed plateaus.
The expected plateau values are therefore
\begin{equation}
    \frac{1}{2N},\frac{2}{2N},\ldots,\frac{2N-1}{2N},1,
\end{equation}
where the final value $1$ corresponds to all coherent components being fully captured by the finite-rank approximation.
\begin{figure*}[t]
    \centering    \includegraphics[width=\textwidth]{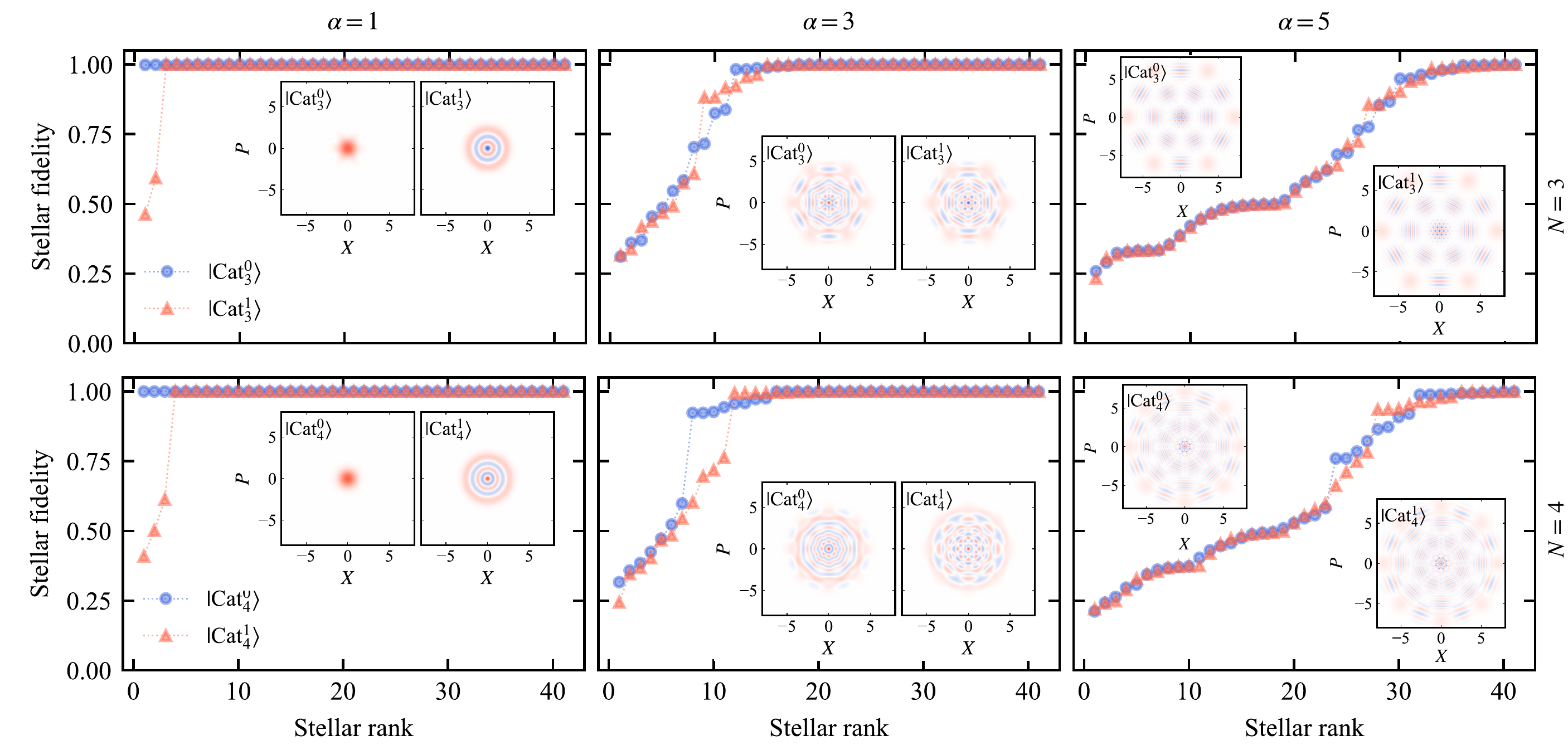}
    \caption{
    Stellar fidelity as a function of stellar rank $k$ for logical cat-code states with $N=3$ and $N=4$, evaluated at $\alpha=1,3,5$.
Blue circles and red triangles denote the two logical codewords $|\mathrm{Cat}_{N}^{0}\rangle$ and $|\mathrm{Cat}_{N}^{1}\rangle$, respectively.
As $\alpha$ increases, the coherent components become more separated in phase space and plateau structures emerge in the stellar-fidelity profiles.
The possible plateau values are determined by the number $2N$ of coherent components, $f_k^\star\approx q/(2N)$.
For $\alpha=5$, the $N=3$ cat code shows more clearly resolved plateaus than the $N=4$ cat code because neighboring coherent components are more widely separated.
    }
\label{fig:stellar_fidelity_cat_N3N4}
\end{figure*}

In practice, not all plateau levels are equally visible.
If several coherent components are captured over overlapping ranges of $k$, adjacent plateaus can merge.
The visibility of the plateau structure is controlled by the distinguishability of neighboring coherent components, characterized by the nearest-neighbor distance
\begin{equation}
d_{\min}
=
2\lvert\alpha\rvert
\sin\left(\frac{\pi}{2N}\right).
\end{equation}
In the large-$N$ limit, this distance scales as
\begin{equation}
d_{\min}
=
\frac{\pi\lvert\alpha\rvert}{N}
\left[1-\frac{\pi^{2}}{24N^{2}}+\mathcal{O}\left(N^{-4}\right)\right]
\approx
\frac{\pi\lvert\alpha\rvert}{N}.
\end{equation}
At fixed $N$, increasing $\lvert\alpha\rvert$ enlarges $d_{\min}$ and makes the coherent components more distinguishable, so the plateau structure becomes easier to resolve.
In contrast, at fixed $|\alpha|$, increasing $N$ reduces $d_{\min}$, causing the capture ranges of different components to overlap more strongly and making the plateaus less pronounced.

This behavior is illustrated in Fig.~\ref{fig:stellar_fidelity_cat_large_alpha}.
For completeness, we show a large-amplitude example with $\alpha=7$, beyond the parameter regime used for the QEC performance analysis in the main text.
The larger amplitude makes the coherent components well separated and hence reveals the plateau structure clearly.
For $N=2$, the cat state contains four coherent components, so the expected plateau values are
\begin{equation}
    \frac14,\quad \frac12,\quad \frac34,\quad 1 .
\end{equation}
The numerical profile exhibits pronounced steps near these values, consistent with the sequential capture of coherent components by the finite-rank approximation.

The same mechanism explains the trends observed in Fig.~\ref{fig:stellar_fidelity_cat_N3N4}.
For $\alpha=5$, plateau structures are visible for both $N=3$ and $N=4$, but they become less sharp as $N$ increases because neighboring coherent components are closer in phase space.
By the same reasoning, an $N=2$ cat state at the same amplitude would exhibit an even clearer plateau structure.
This is precisely what is highlighted by the large-amplitude $N=2$ example in Fig.~\ref{fig:stellar_fidelity_cat_large_alpha}.

This also clarifies why analogous plateau structures are not clearly observed for the GKP states in Fig.~\ref{fig:stellar_profile_wignerfunction_gkp}.
Although GKP codewords can be represented in a coherent-state basis, the coherent amplitudes are fixed by the lattice spacing $\sqrt{\pi/2}$ and the state involves many coherent components.
As a result, the effective component separation is relatively small while the number of components is large, so the discrete component-capture plateaus are washed out in the stellar fidelity profiles.

\section{Additional results on performance of approximate cat codes}
\label{appendix:performance of codewords}

In this Appendix, we present additional numerical results for approximate cat-code states with higher rotational symmetry orders $N=3$ and $N=4$, constructed under finite stellar-rank constraints.
As in the main text, the approximate codewords are obtained by optimizing the stellar fidelity at fixed stellar rank through Gaussian transformations, and their error-correction performance is evaluated using the channel fidelity under optimal recovery.

\begin{figure*}[t]
\centering
\includegraphics[width=\textwidth]{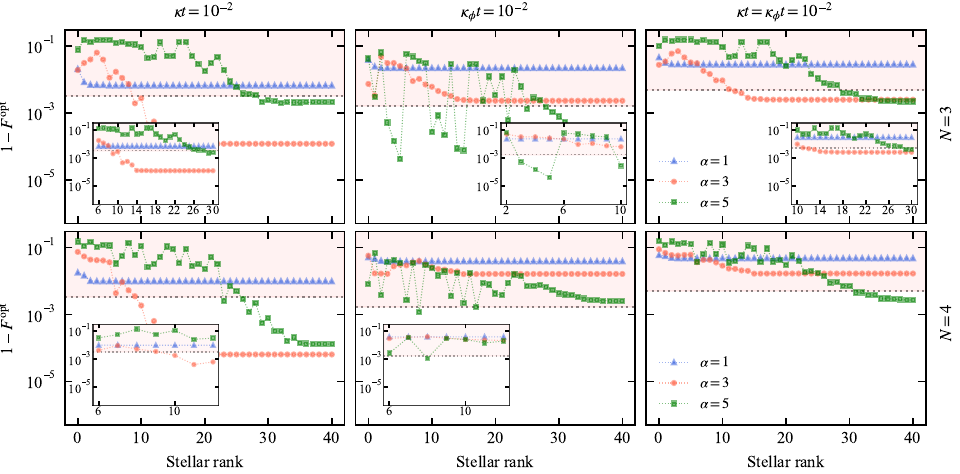}
\caption{
Channel infidelity $(1-F^{\mathrm{opt}})$ as a function of stellar rank $k$ for approximate cat-code states with rotational symmetry $N=3$ and $N=4$.
The upper row shows the $N=3$ cat code, and lower row shows the $N =4$ cat code. Different curves correspond to coherent-state amplitudes $\alpha =1$ (blue triangle), $\alpha=3$ (red circle) and $\alpha = 5$ (green square). From left to right, the columns show photon loss with $\kappa t=10^{-2}$, dephasing with $\kappa_\phi t=10^{-2}$ and combined loss-dephasing noise with $\kappa t=\kappa_\phi t= 10^{-2}$. Markers denote the optimized channel fidelity at selected stellar ranks, while the dotted line are guide to the eye. The horizontal dashed lines indicate the break-even infidelity, and the shade regions mark regimes where the finite-rank approximate codewords perform worse than this threshold.
Larger-amplitude cat states generally require higher stellar rank to reach comparable logical performance, reflecting their increased phase-space separation and greater approximation cost. Insets zoom in on representative stellar-rank interval near the break-even crossings.}
    \label{fig:channel_infidelity_cat_N3N4}
\end{figure*}

Fig.~\ref{fig:stellar_fidelity_cat_N3N4} shows the stellar-fidelity profiles for the higher-order cat codes.
As in the $N=2$ case discussed in the main text, the optimized stellar fidelity generally increases with stellar rank, while larger coherent-state amplitudes require larger stellar-rank resources to accurately reproduce the full rotational structure of the target states.
The higher-order cat codes also exhibit pronounced plateau structures.
These plateaus correspond to intermediate regimes in which the finite-rank approximation resolves only a subset of the $2N$ coherent components.
Increasing the rotational symmetry order therefore makes the approximation progressively more demanding.
For $\alpha=5$, the plateaus are more clearly resolved for $N=3$ than for $N=4$, since neighboring coherent components remain more widely separated at smaller $N$.

Fig.~\ref{fig:channel_infidelity_cat_N3N4} shows the corresponding channel infidelity as a function of stellar rank under photon loss, dephasing, and combined loss-dephasing noise.
Across all noise models, increasing the stellar rank generally improves the channel fidelity until saturation.
Small oscillations at low stellar rank are again associated with the fact that stellar-fidelity optimization is performed at the level of individual target states rather than directly on the encoded subspace relevant to QEC.

The dependence on the cat amplitude is strongly noise dependent.
Under pure dephasing, increasing the amplitude can improve the ideal code performance by separating the coherent components further in phase space, but this advantage appears only once the finite-rank approximation is accurate enough. Thus larger amplitudes may require substantially higher stellar rank before their dephasing-protection advantage become visible. This behavior is evident in the insets, which resolve the stellar-rank intervals near the break-even crossings.

Under photon loss and combined loss-dephasing noise, the trade-off is different: larger amplitudes have broader photon-number support and are therefore more susceptible to loss, while also being harder to approximate at finite stellar rank. Consequently, the stellar rank required to approach or surpass break-even depends sensitively on both the amplitude and the noise channel. 
The comparison between $N=3$ and $N=4$ further confirms the connection between stellar-fidelity plateaus and operational performance: higher rotational symmetry requires resolving a larger number of coherent components, and therefore larger stellar-rank resources are needed before the channel fidelity saturates.

\section{Optimized encodings with fixed stellar rank}
\label{appendix:optimal_states}
\begin{figure}[h]
    \centering
    \includegraphics[width=\columnwidth]{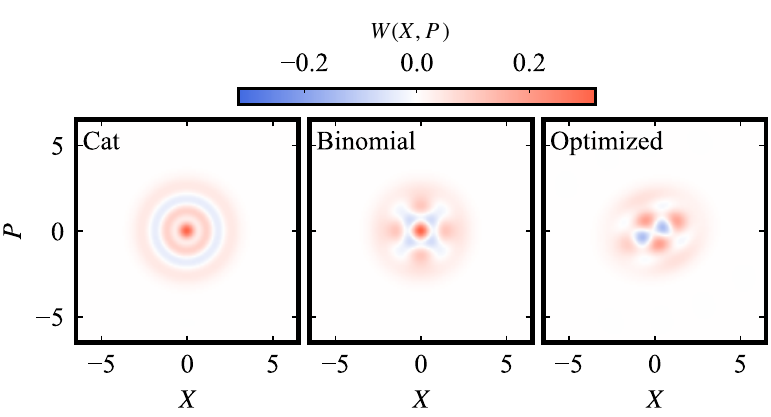}
    \caption{Wigner functions of the maximally mixed encoded states for the three encodings compared at $k=4$ and $\kappa t=10^{-2}$.
    From left to right: the finite-rank approximation of the $N=2$ cat code
    with $\alpha=1.537$, the binomial code in
    Eq.~\eqref{eq:N2_binomial_code}, and the optimized finite-stellar-rank
    encoding.
}
    \label{fig:rank4_wigner_comparison}
\end{figure}
In Sec.~\ref{sec:QEC_optimal_codewords}, we compared three encodings at the
common resource scale $k=4$ and loss rate $\kappa t=10^{-2}$.
Fig.~\ref{fig:rank4_wigner_comparison} shows their maximally mixed encoded
states, providing a basis-independent comparison of the corresponding phase-space structures.

In this Appendix, we present the full set of phase-space structures obtained from the optimization of encodings under fixed stellar-rank constraints.
As described in the main text, for each noise model and error rate we optimize the encoding using the Petz-recovery fidelity as a surrogate objective.

\begin{figure*}[t]
    \centering    \includegraphics[width=\textwidth]{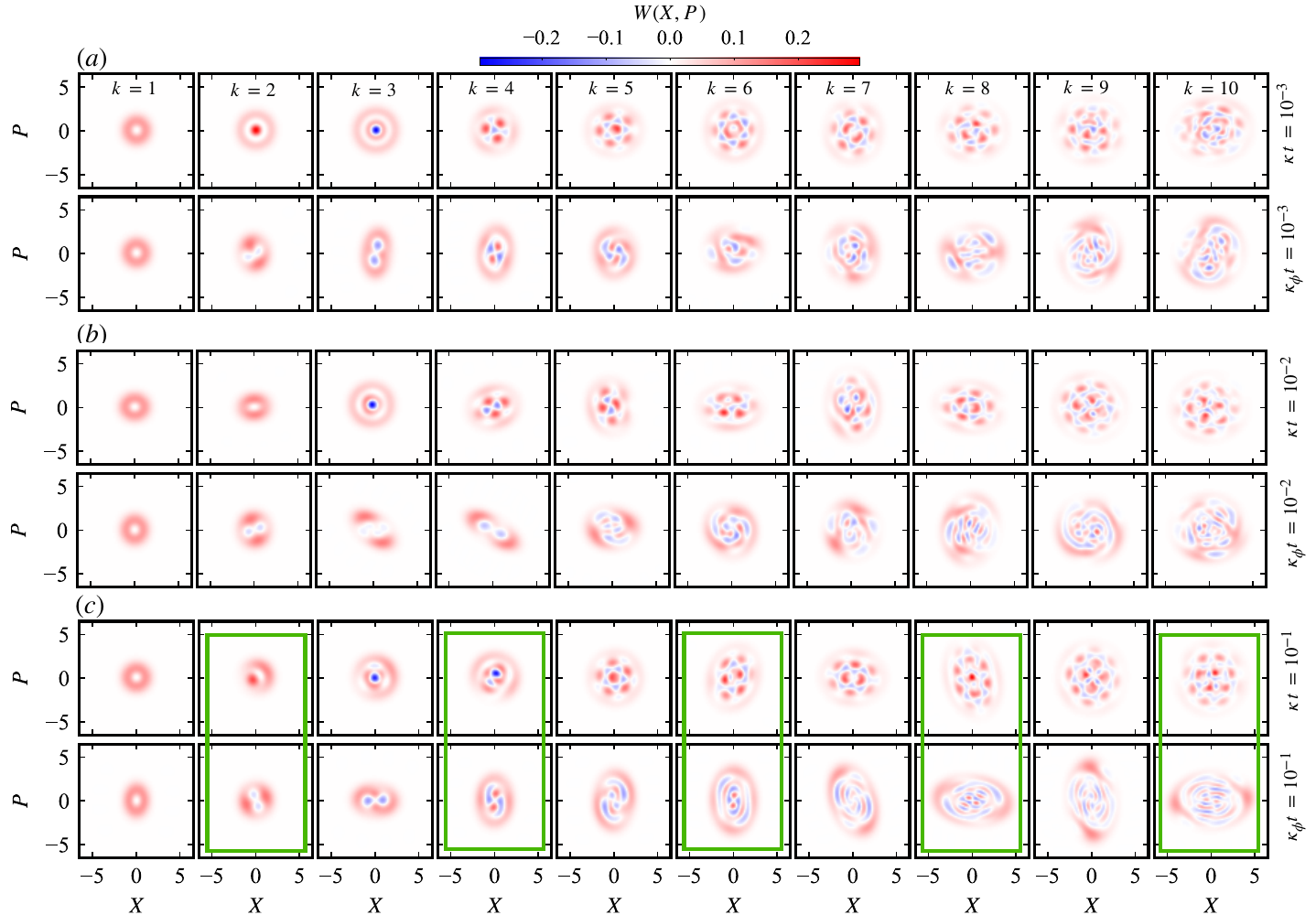}
    \caption{
    Wigner functions of the maximally mixed encoded states obtained from optimized encodings under a fixed stellar-rank constraint.
    Each column corresponds to a different stellar rank, ranging from $k=1$ to $k=10$ from left to right.
    Panels (a), (b), and (c) correspond to increasing noise strengths, $\kappa t,\kappa_\phi t=10^{-3},10^{-2},10^{-1}$, respectively.
    Within each panel, the upper row shows photon-loss noise with rate $\kappa t$, while the lower row shows dephasing noise with rate $\kappa_\phi t$.
    The green frames indicate the representative states shown in Fig.~\ref{fig:entanglement_fid_optimal_states}(c) of the main text.
    The green frames indicate the representative states shown in
    Fig.~\ref{fig:entanglement_fid_optimal_states}(c) of the main text.
    For each noise model and error rate, the encoding is optimized using the
    Petz-recovery fidelity as a surrogate objective.
    As the stellar rank increases, the optimized states develop increasingly structured phase-space patterns: photon loss favors more grid-like structures, while dephasing favors approximate $\pi$-rotation symmetry.}
    \label{fig:optimal_encodings_all}
\end{figure*}

Fig.~\ref{fig:optimal_encodings_all} shows the Wigner functions of the optimized maximally mixed encoded states.
The three blocks correspond to increasing noise strengths, while the upper and
lower rows in each block show the results for photon loss and dephasing,
respectively. The panels framed in green are those displayed in the main text and are
reproduced here for completeness.

The full set of Wigner functions illustrates the systematic evolution of the optimized encodings as a function of stellar rank and noise strength.
As the stellar-rank budget increases, the optimized states develop more structured phase-space features.
The results also show the same qualitative noise-dependent behavior discussed in the main text, with loss-optimized states becoming more grid-like and dephasing-optimized states developing approximate rotational symmetry.

For the smallest error rates considered, the optimized states do not yet display pronounced noise-adapted structures.
In this regime, the noise perturbation is weak, and the channel fidelity is only weakly sensitive to the detailed geometry of the encoded states.
As a result, the optimization has less incentive to develop strongly structured phase-space features.
The characteristic grid-like or rotationally symmetric patterns become more visible as the error rate increases, where adapting the encoding to the dominant noise mechanism has a larger effect on the logical fidelity.

\FloatBarrier
\bibliography{bibliography}

\end{document}